\documentclass[11pt,twoside,a4paper,cmspaper,final,collab]{cms-tdr}
\def\svnVersion{5e0f295-D}\def\svnDate{2026/08/04}\def\cmsCernNoTag{CERN-EP-2026-198}\def\cmsCernDate{\today}\def\cmsMessage{Submitted to the European Physical Journal C}
\begin{document}\cmsNoteHeader{EXO-24-029}

\newcommand{\mt}{\ensuremath{m_{\mathrm{T}}}\xspace}
\newcommand{\MT}{\mt}
\newcommand{\RT}{\ensuremath{R_{\mathrm{T}}}\xspace}
\newcommand{\PZprime}{\ensuremath{{\PZ}^{\prime}}\xspace}
\newcommand{\PAprime}{\ensuremath{{\mathrm{A}}^{\prime}}\xspace}
\newcommand{\mZprime}{\ensuremath{m_{\PZprime}}\xspace}
\newcommand{\sigmaZprime}{\ensuremath{\sigma_{\PZprime}}\xspace}
\newcommand{\wZprime}{\ensuremath{\Gamma_{\PZprime}}\xspace}
\newcommand{\Bdarkexpl}{\ensuremath{\mathcal{B}(\PZprime \to \chi \overline{\chi})}\xspace}
\newcommand{\Bdark}{\ensuremath{\mathcal{B}_{\text{dark}}}\xspace}
\newcommand{\mDark}{\ensuremath{m_{\text{dark}}}\xspace}
\newcommand{\mRhoDark}{\ensuremath{m^{\rho}_{\text{dark}}}\xspace}
\newcommand{\aDark}{\ensuremath{\alpha_{\text{dark}}}\xspace}
\newcommand{\lamDark}{\ensuremath{\Lambda_{\text{dark}}}\xspace}
\newcommand{\rinv}{\ensuremath{r_{\text{inv}}}\xspace}
\newcommand{\Pqdark}{\HepParticle{\upchi}{}{}\xspace}
\newcommand{\Paqdark}{\HepAntiParticle{\upchi}{}{}\xspace}
\newcommand{\PqdarkO}{\HepParticle{\upchi}{1}{}\xspace}
\newcommand{\PqdarkT}{\HepParticle{\upchi}{2}{}\xspace}
\newcommand{\Ppidark}{\ensuremath{\pi_{\text{dark}}}\xspace}
\newcommand{\PpidarkDM}{\ensuremath{\Ppidark^{\mathrm{DM}}}\xspace}
\newcommand{\Prhodark}{\ensuremath{\rho_{\text{dark}}}\xspace}
\newcommand{\PrhodarkDM}{\ensuremath{\Prhodark^{\mathrm{DM}}}\xspace}
\newcommand{\gq}{\ensuremath{g_{\PQq}}\xspace}
\newcommand{\gqdark}{\ensuremath{g_{\Pqdark}}\xspace}
\newcommand{\gu}{\ensuremath{g_{\PQu}}\xspace}
\newcommand{\gtau}{\ensuremath{g_{\PGt}}\xspace}
\newcommand{\brtau}{\ensuremath{\mathcal{B}_{\PGt}}\xspace}
\newcommand{\brtauexpl}{\ensuremath{\mathcal{B}({\pi_{\text{dark}} \to \PGtp \PGtm})}\xspace}
\newcommand{\Nc}{\ensuremath{N_{\text{c}}^\text{dark}}\xspace}
\newcommand{\Nf}{\ensuremath{N_{\text{f}}^\text{dark}}\xspace}
\newcommand{\Nstable}{\ensuremath{N_{\text{stable}}}\xspace}
\newcommand{\Nunstable}{\ensuremath{N_{\text{unstable}}}\xspace}
\newcommand{\widejet}{\ensuremath{\mathrm{J}}\xspace}
\newcommand{\narrowjet}{\ensuremath{\mathrm{j}}\xspace}
\newcommand{\detafull}{\ensuremath{\Delta\eta(\widejet_{1},\widejet_{2})}\xspace}
\newcommand{\deta}{\ensuremath{\Delta\eta}\xspace}
\newcommand{\mindphi}{\ensuremath{\Delta\phi_{\text{min}}}\xspace}
\newcommand{\zjets}{{{\PZ}\text{+jets}}\xspace}
\newcommand{\wjets}{{{\PW}\text{+jets}}\xspace}
\newcommand{\znnjets}{\ensuremath{{\PZ}{(\to\Pgn\Pagn)\text{+jets}}}\xspace}
\newcommand{\zlljets}{\ensuremath{{\PZ}{(\to\Pellp\Pellm)\text{+jets}}}\xspace}
\newcommand{\wlnjets}{\ensuremath{{\PW}{(\to\Pell\Pgn)\text{+jets}}}\xspace}
\newcommand{\Imini}{\ensuremath{I_{\text{mini}}}\xspace}
\newcommand{\Iinter}{\ensuremath{I_{\text{inter}}}\xspace}
\newcommand{\ptsub}[1]{\ensuremath{p_{\mathrm{T},#1}}\xspace}
\newcommand{\ptvecsub}[1]{\ensuremath{\vec{p}_{\mathrm{T},#1}}\xspace}
\newcommand{\mjj}{\ensuremath{m_{\widejet\widejet}}\xspace}
\newcommand{\ptvecjj}{\ensuremath{\ptvecsub{\widejet\widejet}}\xspace}
\newcommand{\ptjj}{\ensuremath{\ptsub{\widejet\widejet}}\xspace}
\newcommand{\phijjmiss}{\ensuremath{\phi_{\widejet\widejet,\text{miss}}}\xspace}
\newcommand{\DRmax}{\ensuremath{\Delta R_{\text{max}}}\xspace}
\newcommand{\Pchhad}{\ensuremath{\mathrm{h}^{\pm}}\xspace}
\newcommand{\Pnhad}{\ensuremath{\mathrm{h}^{0}}\xspace}
\newcommand{\abseta}{\ensuremath{\abs{\eta}}\xspace}
\newcommand{\cdead}{\ensuremath{c_{\text{nonfunctional}}}\xspace}
\newcommand{\DRphispike}{\ensuremath{\Delta R(\narrowjet_{1,2},\cdead)}\xspace}
\newcommand{\nmuons}{\ensuremath{N_{\Pgm}}\xspace}
\newcommand{\neles}{\ensuremath{N_{\Pe}}\xspace}
\newcommand{\svjl}{SVJ\Pell}
\newcommand{\svjtau}{SVJ\PGt}
\newcommand{\epseff}{\ensuremath{\epsilon_{\text{eff}}}\xspace}
\newcommand{\pa}{\ensuremath{\mathrm{p}_{\mathrm{a}}}\xspace}
\newcommand{\pb}{\ensuremath{\mathrm{p}_{\mathrm{b}}}\xspace}
\newcommand{\ja}{\ensuremath{\mathrm{j}_{\mathrm{a}}}\xspace}
\newcommand{\jb}{\ensuremath{\mathrm{j}_{\mathrm{b}}}\xspace}
\newcommand{\Tfeat}{\ensuremath{\mathcal{T}^{(i)}_5}\xspace}
\newcommand{\TfeatGen}{\ensuremath{\mathcal{T}^{(i)}}\xspace}
\newcommand{\Deltai}{\ensuremath{\Delta^{(i)}}\xspace}
\newcommand{\kti}{\ensuremath{\kt^{(i)}}\xspace}
\newcommand{\psii}{\ensuremath{\psi^{(i)}}\xspace}
\newcommand{\pta}{\ensuremath{\ptsub{\mathrm{a}}}\xspace}
\newcommand{\ptb}{\ensuremath{\ptsub{\mathrm{b}}}\xspace}
\newcommand{\phia}{\ensuremath{\phi_{\mathrm{a}}}\xspace}
\newcommand{\phib}{\ensuremath{\phi_{\mathrm{b}}}\xspace}
\newcommand{\zi}{\ensuremath{z^{(i)}}\xspace}
\newcommand{\mi}{\ensuremath{m^{(i)}}\xspace}
\newcommand{\PYTORCH}{\textsc{PyTorch}\xspace}
\newcommand{\PYG}{\PYTORCH \textsc{geometric}\xspace}
\newcommand{\ADAM}{\textsc{Adam}\xspace}
\newcommand{\NAa}{\ensuremath{N_{\text{A}}}\xspace}
\newcommand{\NB}{\ensuremath{N_{\text{B}}}\xspace}
\newcommand{\NC}{\ensuremath{N_{\text{C}}}\xspace}
\newcommand{\ND}{\ensuremath{N_{\text{D}}}\xspace}
\newcommand{\Done}{\ensuremath{D_{1}}\xspace}
\newcommand{\Dtwo}{\ensuremath{D_{2}}\xspace}
\newcommand{\Lbce}{\ensuremath{L_{\text{BCE}}}\xspace}
\newcommand{\Ldisco}{\ensuremath{L_{\text{DisCo}}}\xspace}
\newcommand{\Lclosure}{\ensuremath{L_{\text{closure}}}\xspace}
\newcommand{\epsdiscosub}[1]{\ensuremath{\epsilon_{\text{DisCo},#1}}\xspace}
\newcommand{\epsclosure}{\ensuremath{\epsilon_{\text{closure}}}\xspace}
\newcommand{\nonclosure}{\ensuremath{{C}\hspace{-0.70em}/\kern 0.0em}\xspace}
\newcommand{\LUNDNET}{\textsc{LundNet}\xspace}
\newcommand{\ABCDISCOTEC}{\textsc{ABCDisCoTEC}\xspace}
\newcommand{\MDABCDISCOTEC}{\textsc{MD-ABCDisCoTEC}\xspace}
\newcommand{\EDGECONV}{\texttt{EdgeConv}\xspace}

\newlength\cmsTabSkip\setlength{\cmsTabSkip}{1ex}

\cmsNoteHeader{EXO-24-029}

\title{Search for resonant production of lepton-enriched semivisible jets in proton-proton collisions at \texorpdfstring{$\sqrt{s} = 13\TeV$}{sqrt(s) = 13 TeV}}

\date{\today}

\abstract{
This search targets the resonant production of lepton-enriched semivisible jets (SVJs) from a strongly coupled dark sector, using 138\fbinv of proton-proton collision data collected with the CMS detector at the CERN LHC at $\sqrt{s}=13\TeV$. Two scenarios are investigated: jets enriched in all lepton flavors (\svjl signature) and jets predominantly enriched in tau leptons (\svjtau signature). The analysis focuses on final states in which the missing transverse momentum is aligned with jets containing nonisolated leptons. A dual machine-learning strategy is employed, using a graph neural network for jet identification and a fully connected neural network that combines jet- and event-level information to enhance signal sensitivity and background estimation. The signal models assume a heavy \PZprime mediator with a benchmark coupling of 0.25 to standard model quarks, together with prompt decays of unstable dark hadrons. In the \svjl scenario, mediator masses up to 4.7\TeV are excluded at 95\% confidence level, while masses between 1.8 and 3.5\TeV are excluded in the \svjtau scenario. These results provide the first experimental constraints on lepton-enriched semivisible jets.
}

\hypersetup{%
pdfauthor={CMS Collaboration},%
pdftitle={Search for resonant production of lepton-enriched semivisible jets in proton-proton collisions at 13 TeV},%
pdfsubject={CMS},%
pdfkeywords={CMS, dark sectors, beyond standard model}}

\maketitle 

\section{Introduction}\label{sec:intro}

The existence of dark matter (DM) in the universe is among the most compelling pieces of evidence for physics beyond the standard model (SM)~\cite{Bertone:2004pz,Planck:2018vyg}. However, the fundamental nature of DM remains unknown. A wide range of beyond-the-SM (BSM) scenarios introduce DM candidates that not only account for the observed relic abundance, but also allow for their nongravitational interactions with SM particles. In these scenarios, direct DM production in SM particle collisions is possible.

Collider-based DM searches have predominantly focused on models of weakly interacting massive particles (WIMPs)~\cite{Lee:1977ua,Jungman:1995df}, which are theoretically well motivated and lie within a mass range accessible at the CERN LHC. The WIMP production typically yields clean experimental signatures, characterized by large missing transverse momentum (\ptmiss) that recoils against visible SM particles.
Extensive searches for WIMPs at the LHC have explored a variety of channels, including jets, leptons, or photons~\cite{CMS:2018ffd,CMS:2020ulv,CMS:2021far, Sirunyan:2018owy,CMS:2023dof,ATLAS:2024rlu,Aaboud:2018sfi,ATLAS:2019wdu,ATLAS:2021hza,ATLAS:2020xzu,ATLAS:2021kxv,ATLAS:2024kpy}; top quarks~\cite{Sirunyan:2019gfm,ATLAS:2022znu,CMS:2025ncs}; and Higgs bosons~\cite{Sirunyan:2018gdw,CMS:2019ykj,ATLAS:2021shl}, all typically accompanied by substantial \ptmiss, collectively referred to as ``mono-X'' searches. To date, no significant deviations from the SM predictions have been observed, thereby placing stringent constraints on these models across a substantial portion of their parameter space.

Hidden valley (HV) scenarios~\cite{Strassler:2006im} offer extensions of the SM that can address the nature of DM~\cite{Beauchesne:2018myj,Bernreuther:2019pfb,Kribs:2016cew} and the little hierarchy problem~\cite{Craig:2015pha}. These models predict the existence of a new hidden sector of particles, commonly referred to as the dark sector (DS). The DS particles couple to the SM ones through a mediator that may be massive, weakly coupled, or both, giving rise to nonstandard signatures at the LHC beyond conventional mono-X topologies.

In DS scenarios with a force similar to that of the SM quantum chromodynamics (QCD) \ie, whose particles are charged under $SU(\Nc)$, where \Nc is the number of dark QCD colors, visible jet-like final states can be produced. These objects are produced by dark bound states decaying to SM quarks and may appear at various displacements in the detector, depending on the dark hadron lifetime. Long-lived states lead to displaced signatures, such as emerging or trackless jets~\cite{Bai:2011wy,Schwaller:2015gea,Daci:2015hca}, whereas prompt decays result in dark jets~\cite{Park:2017rfb,Cohen:2020afv} or semivisible jets (SVJs)~\cite{Cohen:2015toa,Cohen:2017pzm,Bernreuther:2019pfb}. In the SVJ scenario, a fraction of the dark states remains stable, yielding a distinctive multijet + \ptmiss signature where the \ptmiss is aligned with one of the jets.

The ATLAS and CMS Collaborations have searched for fully hadronic SVJ signatures~\cite{CMS:2018bvr,CMS:2021dzg,ATLAS:2023swa,ATLAS:2023kao,CMS:2024gxp,CMS:2024bvl,ATLAS:2025kuz,ATLAS:2025bsz,CMS:2025fnr}, without observing significant deviations from the SM. However, recent studies~\cite{Cazzaniga:2022hxl,Beauchesne:2022phk} highlight that existing searches lack sensitivity to scenarios with jets enriched in lepton content.

In this paper, we present the first search for the resonant production of SVJs that are enriched in leptonic content. 
In particular, we focus on two scenarios: the first leading to SVJs enriched in all lepton flavors (\svjl signature), proposed in Ref.~\cite{Cazzaniga:2022hxl}, and the second leading to SVJs enriched in tau leptons (\svjtau signature), as proposed in Ref.~\cite{Beauchesne:2022phk}.

We analyze data from proton-proton ($\Pp\Pp$) collisions delivered by the LHC in 2016--2018 and collected with the CMS detector.
The data sample corresponds to an integrated luminosity of 138\fbinv~\cite{CMS:2021xjt}.
The analysis employs a machine-learning (ML) strategy, utilizing a graph neural network (GNN) for jet identification, or ``tagging'', algorithm to discriminate between signal and background. A fully connected network (DNN) approach is used to estimate the background in the signal region (SR).

{\tolerance=1200
First, selections are applied to event-level kinematic variables to achieve general sensitivity to resonant events where the \pt of one of the jets is aligned with \ptvecmiss. A categorization based on the presence of leptons in jets is performed to further suppress the backgrounds, and capture both the \svjl and \svjtau signatures.
Subsequently, for the first time, an extension of the GNN jet tagger \LUNDNET~\cite{Dreyer:2020brq} is used to distinguish between SM jets and SVJs.
Finally, using the output of the jet tagger, a DNN is employed to extract uncorrelated discriminating features to implement an ``ABCD method'' for estimating the background in the SR~\cite{CMS:2025cvw}. This is the first time such a technique has been extended and employed for a resonant search.\par}

{\tolerance=1200
The SM background is dominated by events where \ptmiss is aligned with one of the jets and one or more leptons are produced within the jets. The background events composed exclusively of jets produced through the strong interaction are referred to as QCD multijet events. In QCD processes, the largest component of \ptmiss arises from the mismeasurement of jet energy. For the top quark-antiquark (\ttbar), \wlnjets, and \znnjets background processes, \ptmiss primarily originates from neutrinos escaping the detector. In the \zlljets process, \ptmiss can result both from neutrinos produced in \PGt decays and from the mismeasurement of jets and leptons. 
Leptons can be produced in jet fragmentation or from vector boson decays in the \ttbar, \wlnjets, and \zlljets processes. The SVJ production cross section is probed as a function of the \PZprime mass, invisible component, dark hadron mass, and branching fraction of the dark hadrons to tau leptons, for the \svjtau signature.\par}

The paper is structured as follows. The signal model is described in Section~\ref{sec:sig_model}. A brief description of the CMS detector is given in Section~\ref{sec:cms}. The simulated signal and background samples are presented in Section~\ref{sec:samples}. The trigger strategy and event reconstruction are discussed in Section~\ref{sec:reco}, followed by the event selection in Section~\ref{sec:evtsel}. Section~\ref{sec:tagger} introduces the SVJ tagging algorithm, whereas Section~\ref{sec:bkg} describes the background estimation method relying on an ML-based ABCD approach with decorrelated discriminants. Systematic uncertainties affecting the signal and background predictions are detailed in Section~\ref{sec:syst}. Finally, the statistical interpretation and results are presented in Section~\ref{sec:results}. The paper is summarized in Section~\ref{sec:summary}. Tabulated results are provided in the HEPData record for this analysis~\cite{HEPData}.

\section{Signal modeling}\label{sec:sig_model}

The signal model is characterized by a QCD-like DS with the following content:
\begin{itemize}
\item Fermionic dark quarks \Pqdark (\PqdarkO, \PqdarkT, \ldots) with \Nf flavors;
\item stable and unstable dark pseudoscalar mesons, \PpidarkDM and \Ppidark (dark pions);
\item stable and unstable dark vector mesons, \PrhodarkDM and \Prhodark;
\item a dark gluon that carries an unbroken $SU(\Nc)$ force, described as ``dark QCD'', with the number of colors \Nc, and confinement scale \lamDark;
\item mediators between the SM sector and the DS, depending on the model.
\end{itemize}

\begin{figure*}[htb!]
\centering
\includegraphics[width=1.0\linewidth]{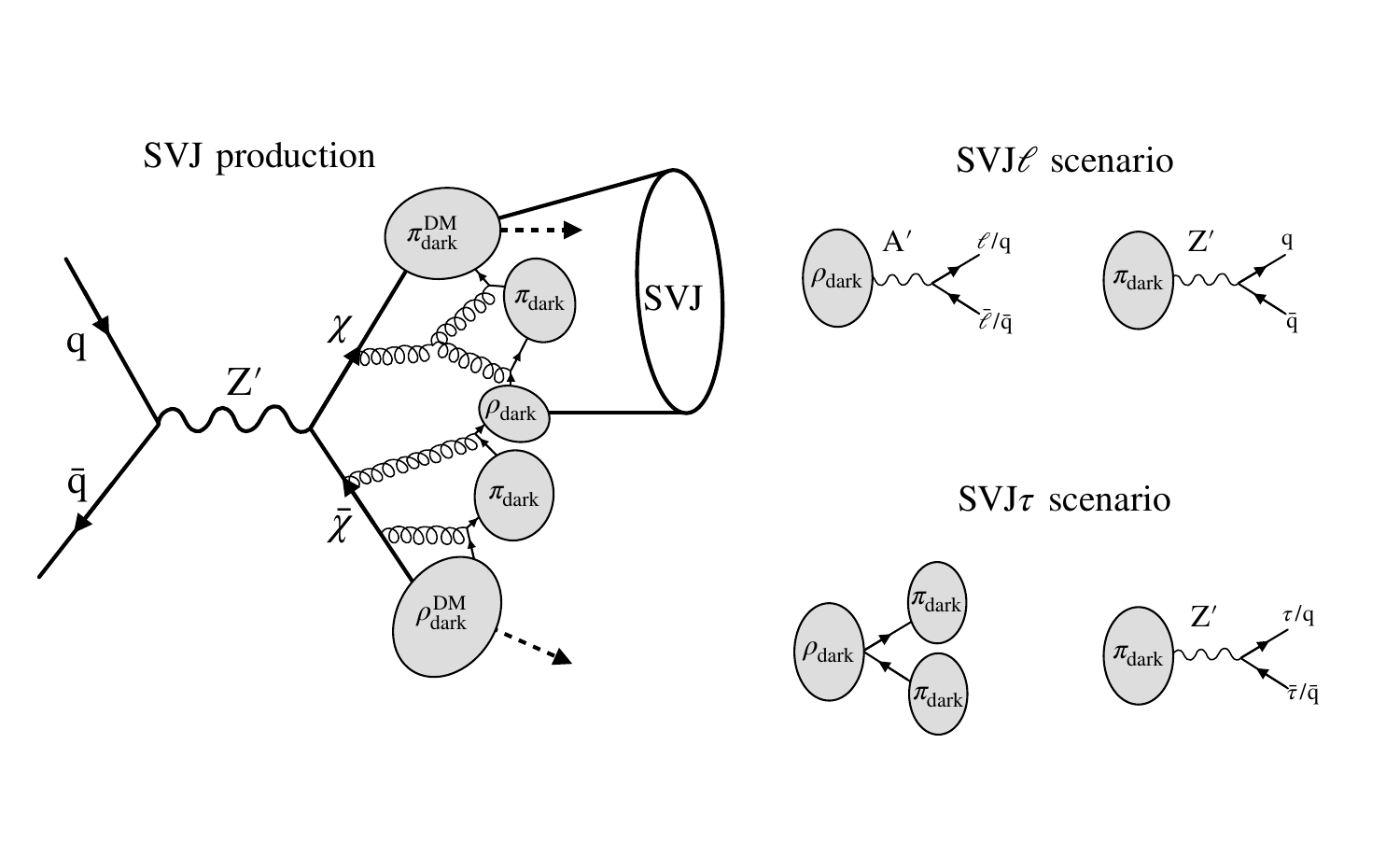}
\caption{Left: diagram of $s$-channel production of SVJs. Right: dark hadrons decay modes in the \svjl and \svjtau scenarios.}
\label{fig:leptons-enr-s-chan-diagram}
\end{figure*}

This analysis targets the resonant production of dark quarks $\cPq\cPaq \to \PZprime \to \Pqdark\Paqdark$.
The final-state dark quarks hadronize via the dark QCD force. Both stable and unstable dark hadrons can be produced. The unstable dark hadrons can decay to SM hadrons and leptons, as shown in Fig.~\ref{fig:leptons-enr-s-chan-diagram}.

In the \svjl model, two mediators connect the SM with the DS: a {\TeVns}-scale leptophobic \PZprime boson that couples both to SM quarks with strength \gq and dark quarks with strength \gqdark, and a lighter dark photon \PAprime that couples both to dark quarks and SM fermions (quarks and leptons). 

Because of the mass hierarchy between \PAprime and \PZprime, the dark vector mesons \Prhodark will decay predominantly via the \PAprime to SM leptons and quarks. The kinetic mixing parameter $\epsilon$ between \PAprime and the SM photon governs the coupling strength of the \PAprime to SM particles. Because of the nature of the \PAprime portal, the vector mesons are expected to exhibit democratic decays to the SM particles. 
 
The final signature is characterized by SVJs containing hadrons along with pairs of oppositely charged leptons (electrons, muons, and taus) produced by the decay of unstable dark bound states. The lifetime of the dark vector mesons is controlled by the parameter $\epsilon$, and prompt decays are possible when the bound state masses are a few {\GeVns} or higher, as discussed in Ref.~\cite{Cazzaniga:2022hxl}, taking into account the constraints from electroweak precision tests (EWPT)~\cite{Curtin:2014cca} and relevant LHC searches~\cite{ATLAS:2019erb,LHCb:2019vmc,CMS:2019buh,CMS:2023hwl}.

In the \svjtau model, a \PZprime vector boson portal, with couplings to the SM quarks and tau leptons, connects the SM and DS particles.
In particular, the \PZprime boson couples to right-handed up-type quarks with strength \gu, to third-generation leptons with strength \gtau, and to dark quarks with strength \gqdark. Within this benchmark, the \Ppidark mesons decay predominantly into the heaviest up-type quark kinematically accessible and into tau leptons. The relative contributions of these two decay modes depend on the relative strengths of the couplings \gu and \gtau.
To probe the impact of unstable dark bound states decaying into tau leptons, an effective parameter $\brtauexpl = \brtau$, which depends on the ratio $\gu/\gtau$, is introduced. 
The final-state signature is characterized by SVJs containing hadrons along with additional electrons, muons, and neutrinos produced by the decay of tau leptons coming from unstable dark hadrons. Since in this case electrons and muons come only from second-stage decays of the dark hadrons, mostly accompanied by the presence of neutrinos carrying away part of the original dark hadron momentum, the resulting transverse momentum (\pt) spectrum of such electrons and muons inside the final-state jets is expected to be softer compared to the \svjl signature. In addition, approximately 65\% of tau leptons decay hadronically, producing charged and neutral hadrons. This leads to a lower fraction of the jet energy being carried by electrons and muons in the \svjtau model compared to the \svjl model, motivating different analyses of the two cases.

\section{The CMS detector}\label{sec:cms}

The CMS apparatus~\cite{CMS:2008xjf,CMS:2023gfb} is a multipurpose, nearly hermetic detector, designed to trigger on~\cite{CMS:2020cmk,CMS:2016ngn,CMS:2024aqx} and identify electrons, muons, photons, and (charged and neutral) hadrons~\cite{CMS:2020uim,CMS:2018rym,CMS:2014pgm}. Its central feature is a superconducting solenoid of 6\unit{m} internal diameter, providing a magnetic field of 3.8\unit{T}. Within the solenoid volume are a silicon pixel and strip tracker, covering $\abseta < 2.5$, a lead tungstate crystal electromagnetic calorimeter (ECAL), covering $\abseta < 3.0$, and a brass and scintillator hadron calorimeter (HCAL), also covering $\abseta < 3.0$, each composed of a barrel and two endcap sections. Forward calorimeters extend the pseudorapidity coverage provided by the barrel and endcap detectors up to $\abseta < 5.2$. Muons are reconstructed using gas-ionization detectors interleaved with the layers of the steel flux-return yoke outside the solenoid, covering $\abseta < 2.4$.
Events of interest are selected using a two-tiered trigger system. The first level, composed of custom hardware processors, uses information from the calorimeters and muon detectors to select events at a rate of around 100\unit{kHz} within a fixed latency of 4\mus~\cite{CMS:2020cmk}. The second level, known as the high-level trigger, consists of a farm of processors running a version of the full event reconstruction software optimized for fast processing, and reduces the event rate to a few kHz before data storage~\cite{CMS:2016ngn,CMS:2024aqx}.
A more detailed description of the CMS detector, together with a definition of the coordinate system used and the relevant kinematic variables, can be found in Refs.~\cite{CMS:2008xjf,CMS:2023gfb}.

\section{Simulation}\label{sec:samples}

{\tolerance=1200
We use simulated Monte Carlo (MC) samples to model the \svjl and \svjtau signals, and SM background processes. For all signal and background samples, \PYTHIA 8.240~\cite{Sjostrand:2014zea} is used to simulate parton showering and hadronization, with the CP5 underlying event tune~\cite{CMS:2022awf} and the NNPDF3.1 next-to-next-to-leading-order (NNLO) parton distribution functions (PDFs)~\cite{Ball:2017nwa}.
Because of changes in detector conditions, including detector upgrades and aging, we use separate samples
for each year of data taking: 2016, 2017, and 2018.
All simulated events are processed through a detailed simulation of the CMS detector using the \GEANTfour software package~\cite{Agostinelli:2002hh}.
Additional interactions within the same or neighboring bunch crossings (pileup) are also included in the simulation.
The simulated samples are corrected using event weights to ensure that the pileup distribution matches the observed distribution in data as closely as possible.\par}

\subsection{Signal simulation}\label{subsec:sng_samples}

Both the \svjl and \svjtau models~\cite{Cazzaniga:2022hxl,Beauchesne:2022phk} have been implemented with \PYTHIA using the HV module~\cite{Carloni:2010tw,Carloni:2011kk}, which facilitates the simulation of a QCD-like hidden sector.
Some of the model parameters are fixed to guarantee a QCD-like behavior of the DS: the number of dark colors $\Nc = 3$, and the number of dark quark flavors $\Nf = 2$. 

The fractions of stable and unstable dark hadrons produced in the dark hadronization process can vary according to the details of the DS. To allow for these variations, an effective invisible fraction is defined following Ref.~\cite{Cohen:2015toa} as: $\rinv = \left\langle \Nstable/(\Nstable+\Nunstable)\right\rangle$. This parameter can take any value between 0 and 1. 

The \Ppidark masses \mDark can differ from the \Prhodark masses \mRhoDark according to the nonperturbative dynamics of the hidden sector, even for mass-degenerate dark quarks. 
Following Ref.~\cite{Albouy:2022cin}, lattice QCD fits have been employed to predict the masses of dark vector mesons from the input ratio $\mDark/\lamDark$.

The DS coupling \aDark is set to run similarly to SM QCD: $\aDark(\lamDark) = \pi/\mathopen{}\left(b_{0}\text{log}\mathopen{}\left(Q/\lamDark\right)\mathclose{}\right)\mathclose{}$ where $b_{0} = (11\Nc - 2\Nf)/6$ and $Q$ is set at the energy scale of the hard scattering. 
Changing the hadronization scale \lamDark will modify the DS coupling strength \aDark, thus influencing the multiplicity and \pt distribution of the dark hadrons.

The \svjl signal model has seven main parameters. Three of these parameters are related to the hidden sector: $\mDark/\lamDark$, \lamDark, and \rinv. 
In the model considered, the \Prhodark vector mesons decay to SM quarks and leptons via the \PAprime portal. To allow such decays, we restrict ourselves to values of $\mDark/\lamDark > 1.5$, thus forbidding the dark vector mesons to decay to the dark pseudoscalar mesons.
The decay of \Prhodark to SM leptons is expected to be democratic with a branching fraction of ${\approx}15\%$ per lepton flavor. The remaining ${\approx}55\%$ of the decays are expected to be to SM quarks, with the decays to up-type quarks in the same generation favored because the electric charge factor enters in the width computation.
The other four parameters are related to the portals of the model, the \PAprime and \PZprime bosons. The couplings of the \PZprime to SM quarks and dark quarks \Pqdark, as well as the mass \mZprime, set the production rate of the process $\cPq\cPaq \to \PZprime \to \Pqdark\Paqdark$ and the \PZprime pole mass, respectively.
We assume that the \PZprime has a universal coupling to SM quarks $\gq = 0.25$ and
a universal coupling to dark quarks $\gqdark$.
The latter coupling is modified from the benchmark value of $\gqdark = 1.0$ given by the LHC DM Working Group~\cite{Albert:2017onk} to $\gqdark = 1.0/\sqrt{\smash[b]{\Nc\Nf}} = 0.4$ to account for the multiple flavors and colors of dark quarks in this signal model, as in Ref.~\cite{CMS:2021dzg}. In particular, the branching fraction $\Bdarkexpl = \Bdark = 46\%$ and mediator width $\wZprime/\mZprime = 5.5\%$, for which the narrow-width approximation holds~\cite{Boveia:2016mrp}.
The last parameter is the effective mixing parameter \epseff, incorporating the nonperturbative DS dynamics information together with the gauge mixing parameter $\epsilon$ and the \PAprime mass. The parameter \epseff governs the decays of the \Prhodark vector mesons to the SM fermions via the \PAprime mediator. Values of \epseff that saturate the EWPT constraints are considered, as in Refs.~\cite{Cazzaniga:2022hxl,Knapen:2021eip}. With such values of \epseff, the dark vector mesons are expected to decay promptly for masses above a few {\GeVns}. Thus, only prompt decays of the unstable dark bound states are simulated.

The \svjtau model also has seven main parameters.
The first three parameters ($\mDark/\lamDark$, \lamDark, and \rinv) are the same as in the \svjl model.
However, in the \svjtau model, only the dark pions can decay back to SM particles, thus the ratio $\mDark/\lamDark$ is set to be less than 1.5. In this way, the dark vector mesons decay only to \Ppidark, which in turn decay to SM quarks and tau leptons via the \PZprime portal.
The last four parameters are related to the portal of the model, the \PZprime boson. 
The mediator, with a pole mass \mZprime, is chosen to have a universal coupling to SM quarks $\gu = 0.25$ and a coupling to dark quarks $\gqdark = 0.6$ as from the benchmark in Ref.~\cite{Beauchesne:2022phk}. In the benchmark considered $\wZprime/\mZprime$ is less than 7\%, for which the narrow-width approximation holds~\cite{Boveia:2016mrp}.
In order to probe the impact of enhanced \PGt production, we scan over the branching fraction \brtau which is equivalent to varying the ratio $\gu/\gtau$. Thus, the cross section for $\cPq\cPaq \to \PZprime \to \Pqdark\Paqdark$ is determined for the chosen values of \gu, \brtau, \gqdark, and \mZprime.
With this setup, \Bdark lies in the range 85--88\%. The production cross sections for the \svjtau model are typically about a factor of two smaller compared to the \svjl model because of the different coupling structures of the corresponding \PZprime portal.
Because of the neutrinos produced in \PGt decays originating from \Ppidark, the effective invisible fraction within the jets is expected to be larger than in the \svjl case. For benchmark couplings with the values assumed in Ref.~\cite{Beauchesne:2022phk}, \Ppidark are expected to decay predominantly to \PQc/\PQb quarks and \PGt. The relative contributions of these two decay modes depend on the relative strengths of the \gu and \gtau couplings.

The signal model parameter values are varied one at a time when generating simulated signal samples. For both \svjl and \svjtau models, \mZprime is varied between 1500 and 5000\GeV in steps of 500\GeV, while the invisible fraction \rinv is varied with values of 0.3, 0.5, and 0.7. 

In the \svjl model, to allow \Prhodark to decay to SM quarks and leptons, values of $\mDark/\lamDark$ equal to 1.6 are considered, for \lamDark values of 5, 10, and 20\GeV.
In the \svjtau model, two benchmark values of 0.8 and 1 are considered for $\mDark/\lamDark$, allowing \Prhodark to decay only to \Ppidark, and \lamDark is varied with values of 10 and 15\GeV. In addition, for the \svjtau model, the branching fraction \brtau is assigned one of three values: 0.3, 0.5, and 0.7. The ranges of the parameters of the signal models considered are summarized in Table~\ref{tab:sigscan}.

\begin{table}[!hbtp]
\topcaption{Parameter ranges considered for the \svjl and \svjtau models. The \Prhodark mass values obtained from lattice QCD fits are rounded to the nearest integers.}
\centering
\begin{tabular}{lcc}
\renewcommand{\arraystretch}{1.2}
Parameter & \multicolumn{2}{c}{Models} \\
 & \svjl & \svjtau \\
\hline 
\mZprime [\TeVns{}] & 1.5--5 & 1.5--5 \\
$\mDark/\lamDark$ & 1.6 & 0.8--1 \\
\lamDark [\GeVns{}] & 5--20 & 10--15 \\
\mDark [\GeVns{}] & 8--32 & 8--15 \\
\mRhoDark [\GeVns{}] & 15--62 & 27--40 \\
\rinv & 0.3--0.7 & 0.3--0.7 \\
\brtau & \NA & 0.3--0.7 \\
\end{tabular}
\label{tab:sigscan}
\end{table}

\subsection{Background simulation}\label{subsec:bkg_samples}

{\tolerance=1200
The \MGvATNLO2.6.5 event generator~\cite{Alwall:2014hca}, employing the MLM matching procedure~\cite{Alwall:2007fs}, is used to simulate the \ttbar, \wlnjets, \znnjets, and \zlljets processes. These simulations use leading order (LO) matrix element calculations, including up to three additional partons for \ttbar and up to four additional partons for the \wlnjets, \znnjets, and \zlljets processes. The samples are normalized using NNLO cross sections for these processes~\cite{Beneke:2011mq,Cacciari:2011hy,Baernreuther:2012ws,Czakon:2012zr,Czakon:2012pz,Czakon:2013goa,Li:2012wna}.
The QCD multijets process is simulated with \PYTHIA 8.240  at LO as a $2\to2$ interaction.
The simulations of SM background processes are compared with the signal to optimize various selection criteria for sensitivity,
to check the agreement with observed data, and to train the neural networks used in the analysis.\par}

\section{Event reconstruction and triggering}\label{sec:reco}

The events recorded with the CMS detector are reconstructed using a particle flow (PF) algorithm~\cite{CMS:2017yfk}, which aims to identify all particles in each event.
The PF algorithm combines information from all subdetector systems in an optimized manner.
Each reconstructed particle, also called PF candidate, is identified as a charged hadron, electron, muon, neutral hadron, or photon.
The anti-\kt algorithm~\cite{Cacciari:2008gp, Cacciari:2011ma} is used to cluster the PF candidates into jets.
The missing transverse momentum vector \ptvecmiss is computed as the negative vector \ptvec sum of the PF candidates in an event;
its magnitude is denoted as \ptmiss~\cite{CMS:2019ctu}.
Electron~\cite{CMS:2020uim} and muon~\cite{CMS:2018rym} candidates are required to pass loose multivariate (MVA) or cutoff-based identification criteria, respectively, and to have $\pt > 10\GeV$; electrons are required to satisfy $\abseta < 2.5$, while muons must satisfy $\abseta < 2.4$.

{\tolerance=1200
The candidate vertex with the highest value of summed physics-object $\pt^2$ is taken to be the primary $\Pp\Pp$ interaction vertex,
where the physics objects used for this determination are the jets and the missing transverse momentum.
For the vertex calculation, the tracks assigned to each vertex serve as input to the jet clustering algorithm,
and the missing transverse momentum is the negative vector \ptvec sum of the resulting jets.
Only charged-particle tracks associated with the primary vertex are considered when reconstructing the final collection of PF candidates.
Rejecting tracks associated with other vertices reduces the pileup impact.\par}

Two types of jets are used in this search. One collection of jets 
is reconstructed using a distance parameter of $R = 0.4$ for the clustering algorithm, and the resulting jets are required to satisfy $\pt > 30\GeV$. We denote these reconstructed objects with a lowercase \narrowjet, and we refer to them as ``narrow jets'' in the following.
The narrow jets are employed when measuring the efficiency of the trigger and to mitigate instrumental backgrounds.
Energy corrections are also applied to the narrow jets, employing an area-based pileup subtraction technique~\cite{Cacciari:2007fd,CMS:2020ebo}.
The effect of the narrow-jet energy corrections is propagated to \ptvecmiss.

{\tolerance=1200
The jet clustering algorithm is also employed with a distance parameter of $R = 0.8$
to produce a second collection of jets, which are then required to have $\pt>200\GeV$. This is the primary jet collection used in the analysis to identify SVJs. Signal jets are expected to have a broader radiation pattern than the SM background jets, due to double parton showering (in the dark and SM sectors). We denote these reconstructed objects with an uppercase \widejet, and we refer to them simply as ``jets'' in the following.
The pileup-per-particle identification (PUPPI) algorithm~\cite{Bertolini:2014bba} is used to mitigate the effect of pileup on the jets.
The PUPPI algorithm uses a local shape variable that reflects the collinear versus soft-diffuse structure in the particle's neighborhood.
With this variable, as well as with event pileup properties and tracking information at the reconstructed-particle level,
the algorithm computes the probability that a given neutral PF candidate results from pileup. The candidate momentum is multiplied by this probability~\cite{CMS:2020ebo}.
The jets are also corrected for nonlinearities in the detector energy response~\cite{Khachatryan:2016kdb}.
In simulated samples, the jet \pt is smeared using measured resolutions to match the observed data.
A set of quality criteria is applied to reject spurious jets arising from instrumental sources~\cite{CMS:2019ctu}.\par}

The data set considered in this paper is collected using triggers that place requirements on the narrow-jet \pt or on the \HT,
which is the scalar \pt sum of all narrow jets with $\pt>30\GeV$ and $\abseta < 3.0$.
The triggers used in 2016 (2017--2018) require a jet with $\pt > 450\,(500)\GeV$ or $\HT > 900\,(1050)\GeV$.
The thresholds were raised in 2017 to compensate for higher instantaneous luminosities.
The efficiency for an event to be selected by any of these triggers is measured in observed data
using a data set collected with an independent trigger that requires a muon with $\pt > 50\GeV$.
The selection requirements that ensure a high trigger efficiency are described in the next section.

\section{Event selection}\label{sec:evtsel}

Events are selected if they contain at least two jets coming from the primary vertex. To ensure a clean event topology, any event where the two leading jets, sorted by \pt, are not both within the pseudorapidity range $\abseta < 2.4$ is discarded. This selection criterion effectively suppresses instrumental backgrounds arising from beam halo or radiation-induced noise---characterized by missing transverse momentum accompanied by substantial energy deposits in the endcap region---while minimally impacting the signal efficiency.

The dijet transverse mass \mt is used as the search variable.
In signal processes, it forms a peak with a kinematic endpoint at \mZprime. It therefore approximately reconstructs the mass of the \PZprime mediator. For background processes, \mt is expected to have a smoothly falling spectrum.
The dijet transverse mass is computed from the four-vector of the dijet system and the \ptvecmiss~\cite{Cohen:2015toa}, and it is defined as:
\begin{equation}\label{eq:mt}
\mt^{2} = \mjj^{2}+2\ptmiss\left[\sqrt{\mjj^{2}+\ptjj^2} - \ptjj\cos(\phijjmiss)\right].
\end{equation}
Here, \mjj is the invariant mass of the system composed of the two highest-\pt jets, \ptvecjj is the magnitude of the vector sum of their transverse momenta, and \phijjmiss is defined as the azimuthal angle between the dijet system and the missing transverse momentum. 

The transverse ratio is defined as $\RT = \ptmiss/\mt$ and used in the selection instead of \ptmiss to identify events with invisible particles.
Since \ptmiss is correlated with \mt, applying a selection directly on \ptmiss alters the shape of the background distribution.
A selection on \RT, instead, efficiently removes $t$-channel QCD events, which have high \mt but low \ptmiss, and thus low \RT, without altering the \mt distribution. 

The remaining $t$-channel QCD events have high \mt values arising
from the large $\detafull = \abs{\eta_{\widejet_{1}}-\eta_{\widejet_{2}}}$ separation between the two highest-\pt jets.
We define the region with $\deta < 1.5$  as the low-\deta region, and the region with $1.5<\deta < 2.2$ as the high-\deta region. Signal events are expected to populate the low-\deta region, and therefore, this requirement is part of the SR definition. The high-\deta region is used to validate the background estimation method, as described in Section~\ref{sec:bkg}. We further define a \deta-extended inclusive selection by relaxing the \deta requirement to $\deta < 2.2$. As explained later, this extended phase space is used to train the SVJ identification algorithm and the DNNs for the background estimation.
The efficiency of the jet-based triggers is measured in observed data in the low-\deta region.
Events with $\MT > 1.5\TeV$ are selected to be in the region where the trigger is maximally efficient, defined as an efficiency of at least 95\% of the plateau value. 
The simulated signal and background events are corrected as a function of \mt to match the measured trigger efficiency for each year of data taking.

Events with anomalously high-\ptmiss values can arise from various reconstruction failures, detector malfunctions, or other noncollision backgrounds. These events are rejected by dedicated filters (\ptmiss \text{filters})~\cite{CMS:2019ctu}.
The main source of instrumental background is the mismeasurement of jet energy caused by nonfunctional ECAL readout channels (\cdead).
We employ an additional filter, optimized for the kinematic phase space of this search, to identify the nonfunctional channels as done in~\cite{CMS:2021dzg}.
An unphysical contribution from misreconstructed jets, which are found to produce events with artificially high \MT values,
is rejected by vetoing events where the leading narrow jet has a photon energy fraction $f_{\PGg}>0.7$ and $\pt>1\TeV$, as in Ref.~\cite{CMS:2021dzg}.
In the later portion of the 2018 data-taking period, corresponding to an integrated luminosity of 38.7\fbinv, a section of the HCAL was not active.
Since this effect was not included in the simulation, we reject events in which any narrow jet falls into the inactive section ($-3.05 < \eta < -1.35$, $-1.62 < \phi < -0.82$).

The minimum angle between the direction of each of the two highest-\pt jets and \ptvecmiss is defined as \mindphi.
Events with $\mindphi < 0.8$ are selected because the missing transverse momentum aligns with the jets in signal events.
The signal efficiency after the final selection in this extended phase space is approximately 20\% for a benchmark \svjl model point (BP-\svjl) with $\mZprime = 3\TeV$, $\rinv = 0.3$, and $\mDark = 16\GeV$. A similar efficiency is found for a benchmark \svjtau model point (BP-\svjtau) with $\mZprime = 3\TeV$, $\rinv = 0.3$, $\mDark = 8\GeV$, and $\brtau = 0.3$. After these selections, a reduced fraction of SM background events are retained: only $10^{-7}$ of the QCD background, about $10^{-4}$ of both the \ttbar and \znnjets backgrounds, and roughly $5\times10^{-5}$ of both the \wlnjets and \zlljets backgrounds.

{\tolerance=1200
Both the \svjl and \svjtau signatures are characterized by final-state electrons and muons in the jets. Because of the differences in the decay modes of the dark hadrons in the two types of signals, the resulting electron and muon \pt spectra are expected to be different, necessitating different strategies to capture both signatures.
For this reason, for the inclusive low-\deta, high-\deta, and \deta-extended regions, events are selected depending on the lepton (electron and muon) multiplicity.
A lepton isolation requirement is applied to define sensitive categories that capture the different features of electrons and muons in the \svjl and \svjtau signals while suppressing the backgrounds. The lepton isolation is quantified using the mini-isolation variable~\cite{Rehermann:2010vq,CMS:2021dzg} to select electrons and muons with high \pt inside jets:
\ifthenelse{\boolean{cms@external}}
{
\begin{multline}\label{eq:mini_iso}
\Imini(\Pell) = \frac{1}{\ptsub{\Pell}}\sum_{\DRmax}\Bigg(\ptsub{\Pchhad} \\
+ \max\Bigg[0,\ \sum_{\DRmax}\ptsub{\PGg}+\ptsub{\Pnhad}-\tfrac{1}{2}\ptsub{\text{pileup}}\Bigg]\Bigg),
\end{multline}
}
{
\begin{equation}\label{eq:mini_iso}
\Imini(\Pell) = \frac{1}{\ptsub{\Pell}}\sum_{\DRmax}\Bigg(\ptsub{\Pchhad}+\max\Bigg[0,\sum_{\DRmax}\ptsub{\PGg}+\ptsub{\Pnhad}-\frac{1}{2}\ptsub{\text{pileup}}\Bigg]\Bigg)
\end{equation}
}
where the different particle candidates are denoted by \Pell for the charged lepton, \Pchhad for charged hadrons, \PGg for photons, and \Pnhad for neutral hadrons.
The label ``pileup'' in the last term denotes charged hadrons that do not originate from the primary vertex and are used to estimate additional energy coming from neutral candidates whose vertex cannot be determined.
All the sums consider candidates within a cone of $\Delta R = \sqrt{\smash[b]{(\Delta\eta)^2+(\Delta\phi)^2}}\allowbreak < \DRmax$,
where \DRmax is 0.2 for $\ptsub{\Pell} < 50\GeV$, $10\GeV/\ptsub{\Pell}$ for $50 < \ptsub{\Pell} < 200\GeV$, and 0.05 for $\ptsub{\Pell} > 200\GeV$.
The radius of the cone decreases with increasing charged-lepton \pt to account for the effect of the Lorentz boost of the lepton's parent particle;
a larger boost leads to more collimated decay products.
\par}

\begin{figure*}[htb!]
    \centering
    \includegraphics[width=0.48\linewidth]{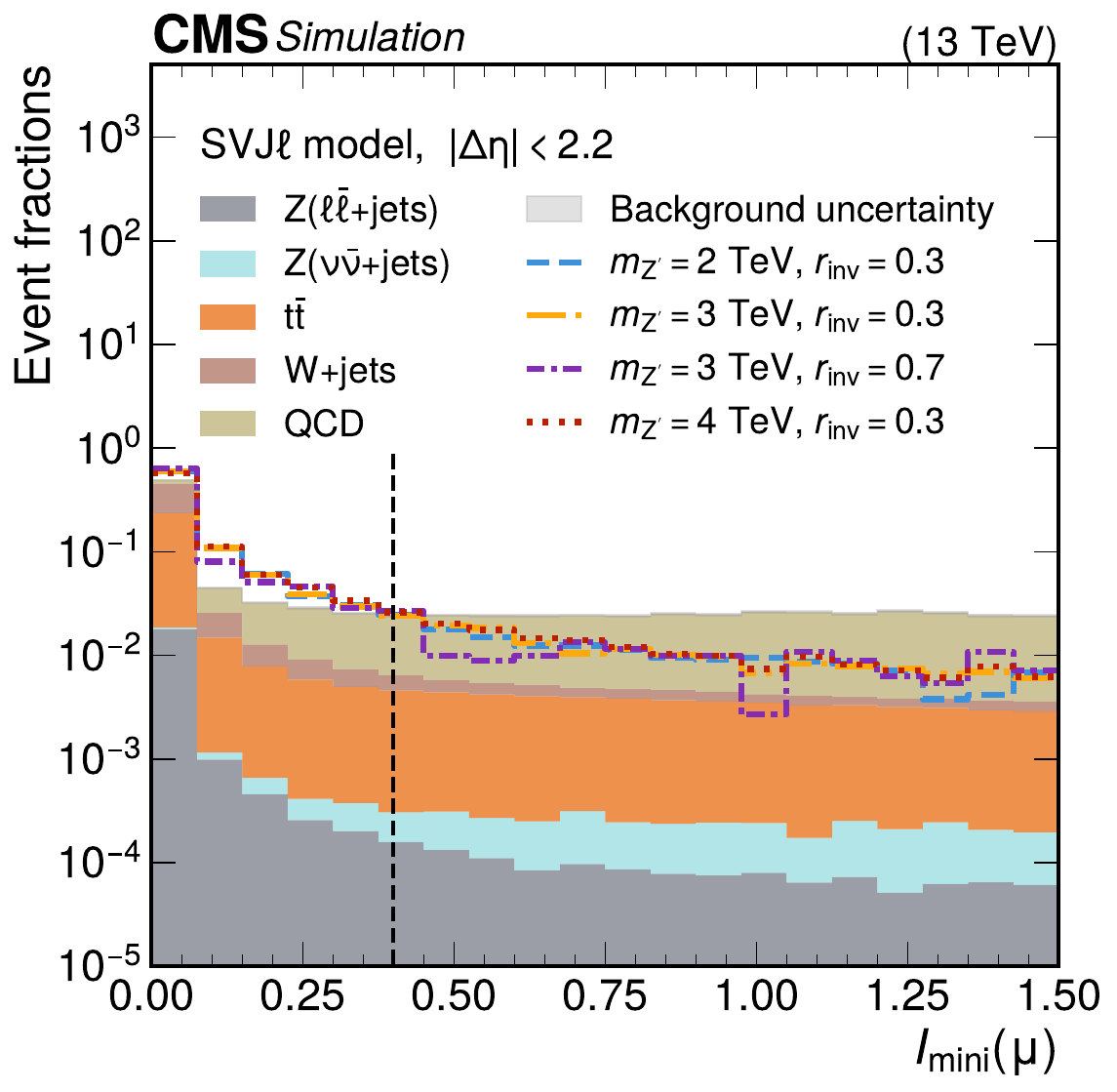}
    \includegraphics[width=0.48\linewidth]{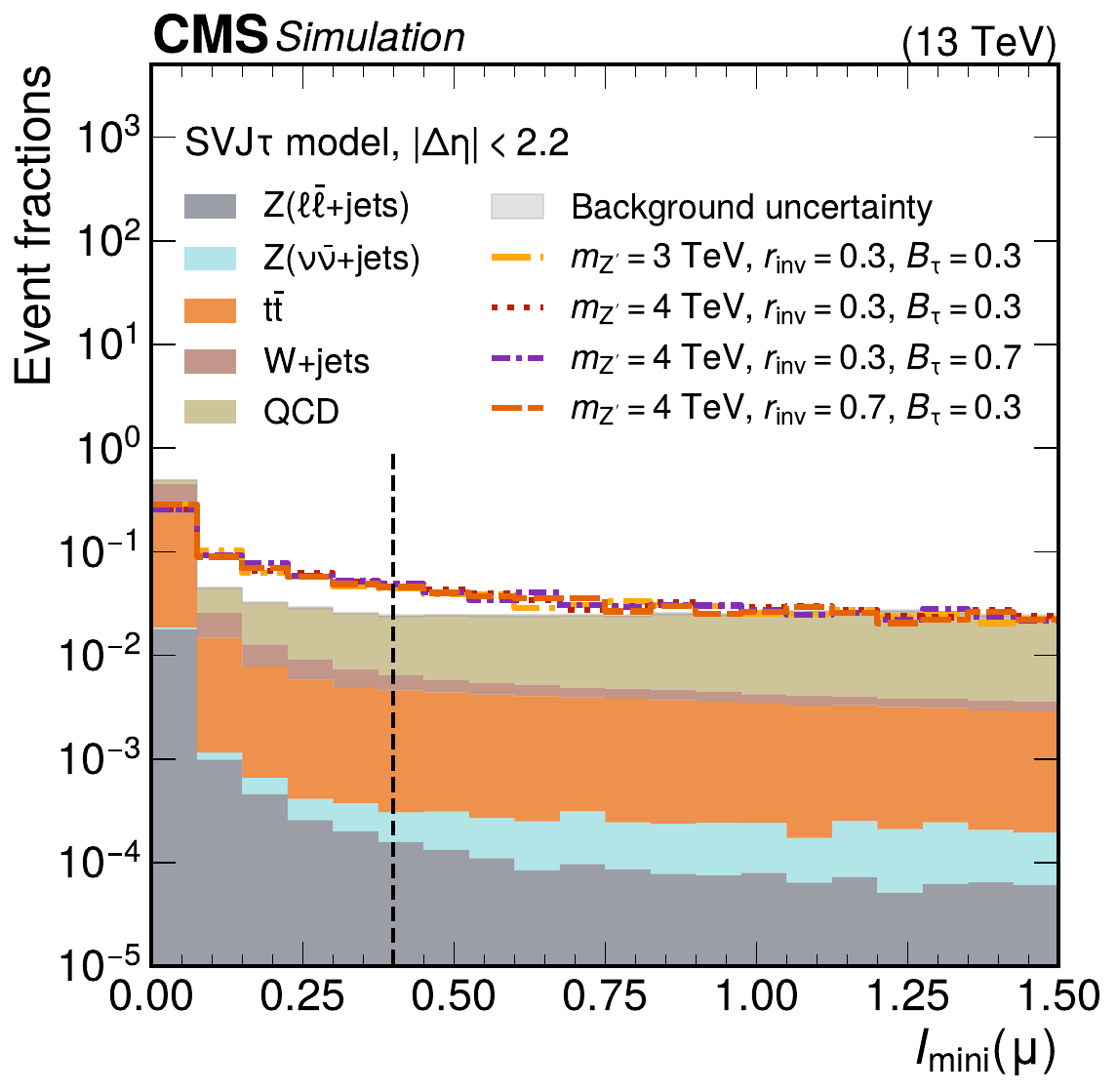}
    \caption{Distribution of $\Imini(\PGm)$ in the \deta-extended region after applying all selection requirements (except for \Imini itself) for simulated background processes and various models of \svjl with $\mDark=16\GeV$ (left) and \svjtau with $\mDark=8\GeV$ (right). The dashed vertical lines indicate the selection requirement for isolated leptons. The sum of the background contributions as well as each individual signal process are normalized to unity.}
    \label{fig:mini_isolation}
\end{figure*}

We define leptons to be isolated if $\Imini(\Pell) < 0.4$. Events with no isolated leptons are assigned to a ``0-lepton'' category, and the rest to a ``multilepton'' one.
As implied by Eq.~\ref{eq:mini_iso}, leptons with high \pt are expected to have smaller isolation cones, and thus smaller $\Imini(\Pell)$ values, whereas leptons with lower \pt are expected to have larger isolation cones and, consequently, larger $\Imini(\Pell)$ values. As shown in Fig.~\ref{fig:mini_isolation}, this categorization provides sensitivity both to signals where the leptons have sufficiently high \pt to pass the mini-isolation requirement (thus entering the multilepton category), as typically expected for leptons from \svjl signals, and to signals where the leptons are softer and accompanied by higher hadronic activity (thus possibly failing the isolation requirement and populating both categories), as expected for leptons from \svjtau signals. For the \svjl model, the multilepton category in the low-\deta region is used for the final signal inference, since most signal events are expected to have leptons passing the mini-isolation requirement. While, for the \svjtau, both the low-\deta 0-lepton and multilepton categories are used in the signal extraction, as a sizable fraction of signal events populate both categories.

{\tolerance=1200
Applying the multilepton requirement retains about 85\% of the BP-\svjl signal model, while rejecting approximately 80\% of the overall background, most notably further removing ${\approx}95\%$ of the remaining QCD contribution. In this multilepton category, the main background is primarily composed of ${\approx}45\%$ \wjets and ${\approx}38\%$ \ttbar events, with QCD and \zjets accounting for smaller fractions. For the \svjtau signal, only part of the signal is retained in the multilepton category, around 35\% for the BP-\svjtau model. In the 0-lepton category, the QCD component dominates the background, since the veto on isolated electrons and muons suppresses mainly the \ttbar and \PW/\PZ + jets contributions.\par}

Both the 0-lepton and multilepton categories in the \deta-extended region are used to validate the agreement between the data and the simulation for the main kinematic features and to train the dedicated \svjl and \svjtau identification algorithms and the DNNs used for background estimation.

The requirements of the inclusive selection and the lepton isolation are summarized in Table~\ref{tab:sel}.

\begin{table*}[!hbt]
   \topcaption{Summary of the inclusive selection and categorization.}
   \centering
   \renewcommand{\arraystretch}{1.5}
   \begin{tabular}{c p{3cm}}  
   \multicolumn{1}{c}{{Inclusive selection}} \\
   \hline
   $\pt(\widejet_{1,2}) > 200\GeV$, $\eta(\widejet_{1,2}) < 2.4$ \\
   \RT  $> 0.15$ \\
   \deta  $< 1.5$ \\
   \mt  $> 1.5\TeV$ \\
   \ptmiss  \text{filters} \\
   \DRphispike  $> 0.1$ \\
   $\text{veto}~f_{\PGg}(\narrowjet_{1}) > 0.7$ \& $\pt(\narrowjet_{1}) > 1.0\TeV$ \\
   $\text{veto}~{-}3.05 < \eta_{\narrowjet} < {-}1.35$  \& ${-}1.62 < \phi_{\narrowjet} < {-}0.82$~~(part of 2018 data only) \\
   \mindphi $< 0.8$ \\
   \multicolumn{1}{c}{{Categorization}}\\
   \hline
   $\pt(\Pe) > 10\GeV$, $\abs{\eta(\Pe)} < 2.5$, loose MVA ID, $\Imini(\Pe) < 0.4$  \\
   $\pt(\PGm) > 10\GeV$, $\abs{\eta(\PGm)} < 2.4$, cutoff-based ID, $\Imini(\PGm) < 0.4$  \\
   0-lepton category: $\neles + \nmuons = 0$ \\ 
   multilepton category: $\neles + \nmuons > 0$ \\
   \end{tabular}
   \label{tab:sel}
\end{table*}

\section{Semivisible-jet tagging}\label{sec:tagger}

The selection requirements described in Section~\ref{sec:evtsel} rely on event-level quantities or basic kinematic properties of the reconstructed jets, leptons, and missing transverse momentum. To further increase the sensitivity, we exploit the intrinsic differences between SVJs and SM jets---arising from their distinct underlying dynamics---to design a dedicated jet-tagging algorithm.

\subsection{Lund tree representation}

We employ \LUNDNET~\cite{Dreyer:2020brq} as a jet tagging method. The \LUNDNET algorithm relies on a graph neural network (GNN) architecture to efficiently tag different types of jet-like structures originating from SM or BSM partons, exploiting their different radiation patterns.
The jet representation employed, called Lund tree, introduces a strong inductive physics-based bias in the GNN, which aims to encode the parton showering and hadronization mechanisms underlying jet formation. 
This aspect enables high-performance tagging of Lorentz-boosted objects, as well as of possible BSM jets with complex substructure, such as SVJs. 

The Lund tree is the generalization of the Lund jet plane~\cite{Dreyer:2018nbf}, which is a two-dimensional representation of the jet emissions pattern. It is built as a binary tree that encodes all the Lund planes of the jet, thus capturing its full radiation pattern. As extensively discussed in Ref.~\cite{Dreyer:2018nbf}, the Lund jet planes can capture the underlying physics processes that lead to the formation of a jet, such as parton showering and hadronization. In the context of dark jets and SVJs, it has been shown~\cite{Cohen:2023mya} that the Lund jet plane can capture new physics scales associated with a strongly coupled hidden sector, such as \mDark and \lamDark, as well as different showering and hadronization mechanisms. Thus, the Lund tree is an ideal tool for discovering jet-like physics signals originating from a new dark sector.

{\tolerance=1200
The Lund tree is obtained by reclustering jet constituents with the Cambridge--Aachen (CA) algorithm~\cite{Bentvelsen:1998ug}. The CA clustering sequentially identifies and combines the pair of particles \pa and \pb closest in $\Delta^2$ into a pseudojet, where $\Delta^2$ is defined as $(y_{\mathrm{a}} - y_{\mathrm{b}})^2 + (\phia - \phib)^2$, $y$ is the rapidity of the particle, and $\phi$ is its azimuthal angle with respect to the jet axis.\par}

At each step of the CA algorithm, the Lund tree is updated with a new node representing a pseudojet. At clustering step $i$,
we label the pseudojet of the previous clustering step with the highest transverse momentum, denoted by \ja, as the `emitter', and the pseudojet with the lowest transverse momentum, denoted by \jb, as the `emission'. To each node of the Lund tree ($i$), we associate a set of five features \Tfeat, as introduced in Ref.~\cite{Dreyer:2018nbf}, that encodes the kinematic information of the splitting $\ja \to \jb$: $\Tfeat = \{ \Deltai, \kti, \zi, \mi, \psii \}$. At each clustering step $i$, $\Delta$ is the angular separation between \ja and \jb; \kt is $\ptb \Delta$; $z$ is the transverse momentum fraction $\ptb/(\pta+\ptb)$; $m$ is the invariant mass of the pair \ja and \jb; and $\psi$ is defined as $\mathrm{tan}^{-1}(\Delta/(\phia-\phib))$.
We further supplement \Tfeat by adding the pseudojet energy fractions carried by charged and neutral hadrons, photons, electrons, and muons. This approach also leverages particle ID information in the Lund tree, which is not accounted for in Ref.~\cite{Dreyer:2018nbf}. This is particularly relevant for tagging the \svjl and \svjtau signals, where an enhanced presence of leptons is expected within the jets.
The energy fractions at step $i$ are defined as $f = \sum_{\mathrm{p}} E_{\mathrm{p}} / \sum_{\mathrm{j}} E_{\mathrm{j}}$, where the sum over $\mathrm{j}$ considers the constituents of the pseudojet formed at step $i$, whereas the sum over $\mathrm{p}$ considers the constituents of a given particle type: charged and neutral hadrons, photons, electrons, and muons.

\begin{figure*}[htb!]
    \centering
    \includegraphics[width=0.99\linewidth]{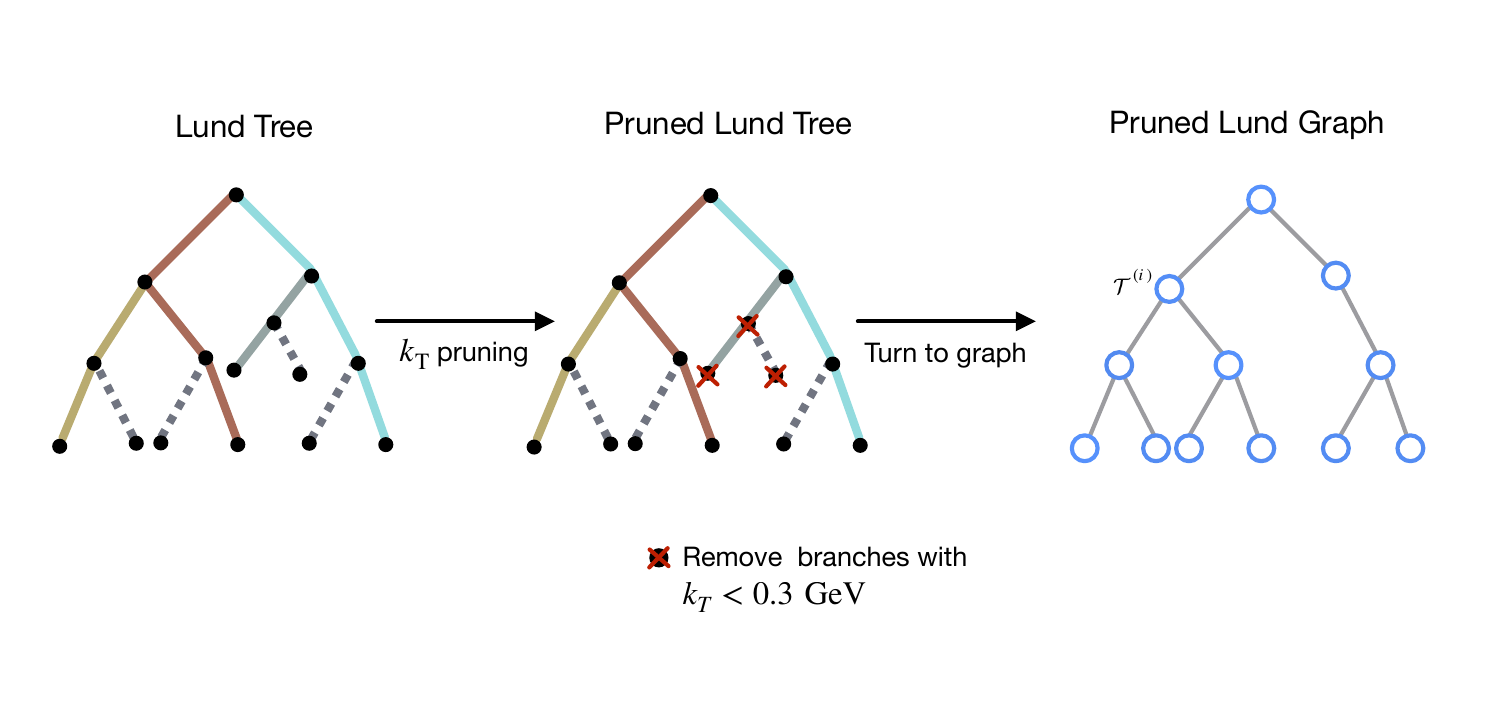}
    \caption{Illustration of the Lund tree before and after pruning, and its conversion into a pruned graph fed to \LUNDNET. The edge colors indicate different Lund planes, with dashed edges indicating further Lund planes that are not fully shown. Each node of the graph has a set of features \TfeatGen associated, representing the kinematic information about the splittings encoded in the Lund tree.
    }
    \label{fig:lund-graph-and-net}
\end{figure*}

{\tolerance=1200
As discussed in Ref.~\cite{Dreyer:2018nbf}, the part of the Lund jet plane that is more dependent on the hadronization model, and thus on nonperturbative effects, is the low-\kt region.
To mitigate such effects, we adopt a \kt-pruning technique, in which all the emissions with $\kt < \kt^{\text{min}}$ are removed from the Lund tree. 
We choose a threshold $\kt^{\text{min}}=0.3\GeV$, which is around the QCD confinement scale. 
This pruning, which removes low-\kt radiation, makes the network resilient against nonperturbative effects, namely the modeling of hadronization in the simulated samples used for training. The \kt-pruning procedure is exemplified in Fig.~\ref{fig:lund-graph-and-net}.\par}

\subsection{Model architecture, training, and performance}

The pruned Lund tree described above is converted into a graph representation, consisting of a node feature matrix and an edge tensor, and used as input to a GNN for jet tagging. The edges of the Lund graph are bidirectional and are defined by the parent-child relationship from the Lund declustering history.
The GNN architecture employed for \LUNDNET is based on the \EDGECONV operation~\cite{Wang:2019}, a graph convolutional layer that aggregates information from a node's neighbors. Because of the tree structure of the Lund graph, aggregating information from neighbors is faster than in a generic fully connected graph, since there are at most 3 edges per node, thereby reducing computational cost. The \LUNDNET therefore improves both the speed both of the training and the inference compared to a GNN using fully connected graphs. This improvement is enhanced by the use of higher-level kinematic inputs, which means the \LUNDNET models converge to a good solution in significantly fewer training epochs.

{\tolerance=1200
The architecture of the \LUNDNET model employed follows Ref.~\cite{Dreyer:2018nbf}. It is built by stacking six \EDGECONV blocks to form a deep graph network. The sizes of the multilayer perceptrons (MLPs) in the six \EDGECONV blocks are $(32,32)$, $(32,32)$, $(64,64)$, $(64,64)$, $(128,128)$, and $(128,128)$, respectively. The outputs from these \EDGECONV blocks are concatenated per node and further processed by another MLP with dimensions $(448,384)$ to better aggregate features learned at different stages. An averaging step is applied to read out information from all nodes in the graph.
After that, the model uses another layer of size $(384,128)$, adds a dropout to prevent overfitting, and finishes with a final layer that makes the classification decision, sized $(128,1)$.\par}

We implement the \LUNDNET models with \PYG~2.3.1 using the \PYTORCH~2.0.1 backend~\cite{Paszke:2019}. The training is performed on GPUs with a batch size of 256. The \ADAM optimizer~\cite{Kingma:2014vow} is used to minimize the cross-entropy loss for the jet classification task.
The training inputs are the Lund graphs of the two highest \pt jets from simulated signals and background MC samples. A \LUNDNET is trained for each category of the analysis in the \deta-extended inclusive selection, and separately for the \svjl and \svjtau signals. Specifically, for the \svjl model, a \LUNDNET is trained in the multilepton category, whereas for the \svjtau model, two \LUNDNET models are trained: one in the 0-lepton category and another in the multilepton category.
Each \LUNDNET model is trained on a mixture of all signal hypotheses from the scan described in Section~\ref{subsec:sng_samples}, with each signal sample assigned equal weight. The background jets are drawn from an equal mixture of SM background processes.

\begin{figure*}[htb!]
    \centering
    \includegraphics[width=0.48\linewidth]{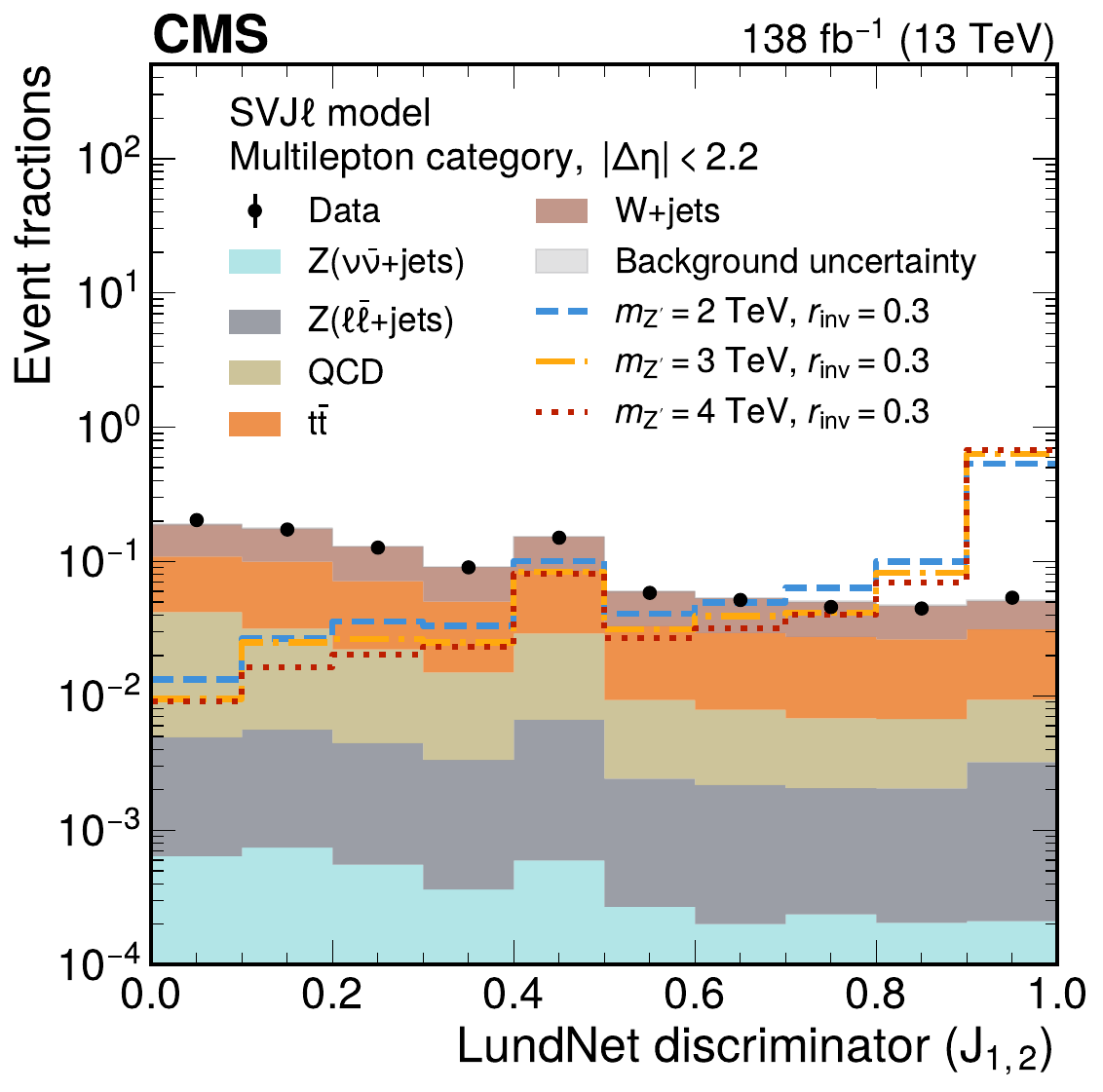}
    \includegraphics[width=0.48\linewidth]{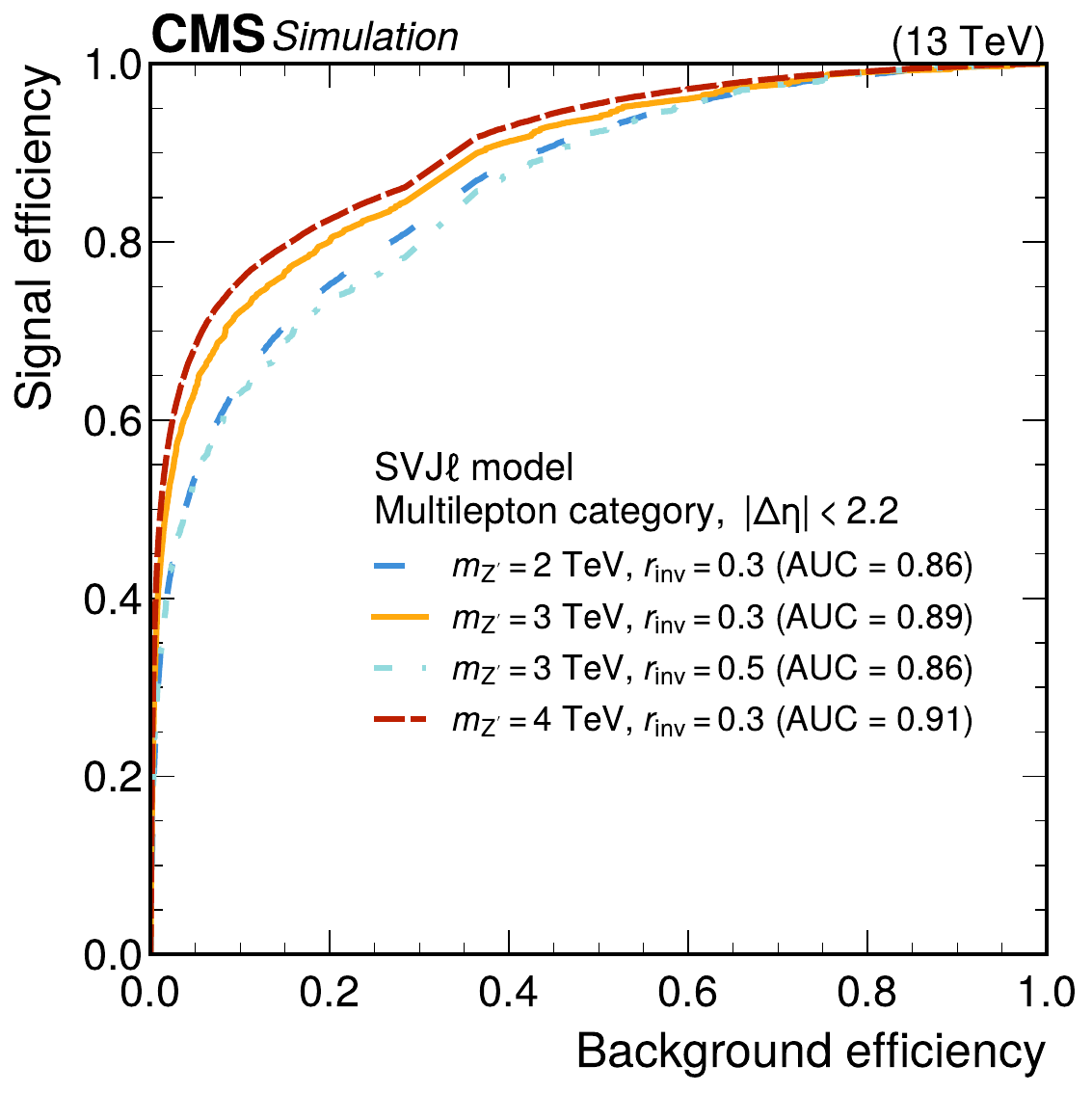}
    \caption{Left: \LUNDNET jet tagger score for the two highest \pt jets in the multilepton category (\deta-extended region) for different \svjl signal models (with $\mDark=16\GeV$), simulated backgrounds, and data. The sum of the background contributions, data as well as each individual signal process are normalized to unity. Statistical uncertainties in the data are shown with vertical bars on the marker. Right: the ROC curves presenting the possible working points of \LUNDNET in the background efficiency versus signal efficiency plane from simulations for different \svjl signal models. The AUC is computed as the area under the ROC curve for a given signal model against the total background.}
    \label{fig:lundet_performance_svjl}
\end{figure*}

The trained \LUNDNET models exhibit strong and uniform rejection of jets from all SM background processes, while correctly selecting a significant fraction of signal jets. The performance of the \LUNDNET discriminator for the \svjl signal is illustrated in Fig.~\ref{fig:lundet_performance_svjl}: the left panel shows the score distribution, while the right panel shows the receiver operating characteristic (ROC) curves, presenting the possible working points of \LUNDNET in the background efficiency-signal efficiency plane for different \svjl signal models. The global discrimination power of the \LUNDNET discriminator is quantified by computing the area under the ROC curve (AUC), evaluated for each signal model against the total background. The \LUNDNET scores for signal and background both exhibit a peak between 0.4 and 0.5. This arises from the \kt-pruning procedure, which, while improving robustness by removing poorly modeled information from the Lund tree, can lead to very shallow trees. In such cases, distinguishing between signal and background jets becomes challenging. This feature is common to all the three \LUNDNET models trained, and is reflected as a kink in the ROC curves. The discrimination power if the \LUNDNET tagger increases at higher \mZprime and lower \rinv values due to the harder momentum spectrum and lower information loss due to undetected particles, respectively.

\begin{figure*}[htb!]
    \centering
    \includegraphics[width=0.48\linewidth]{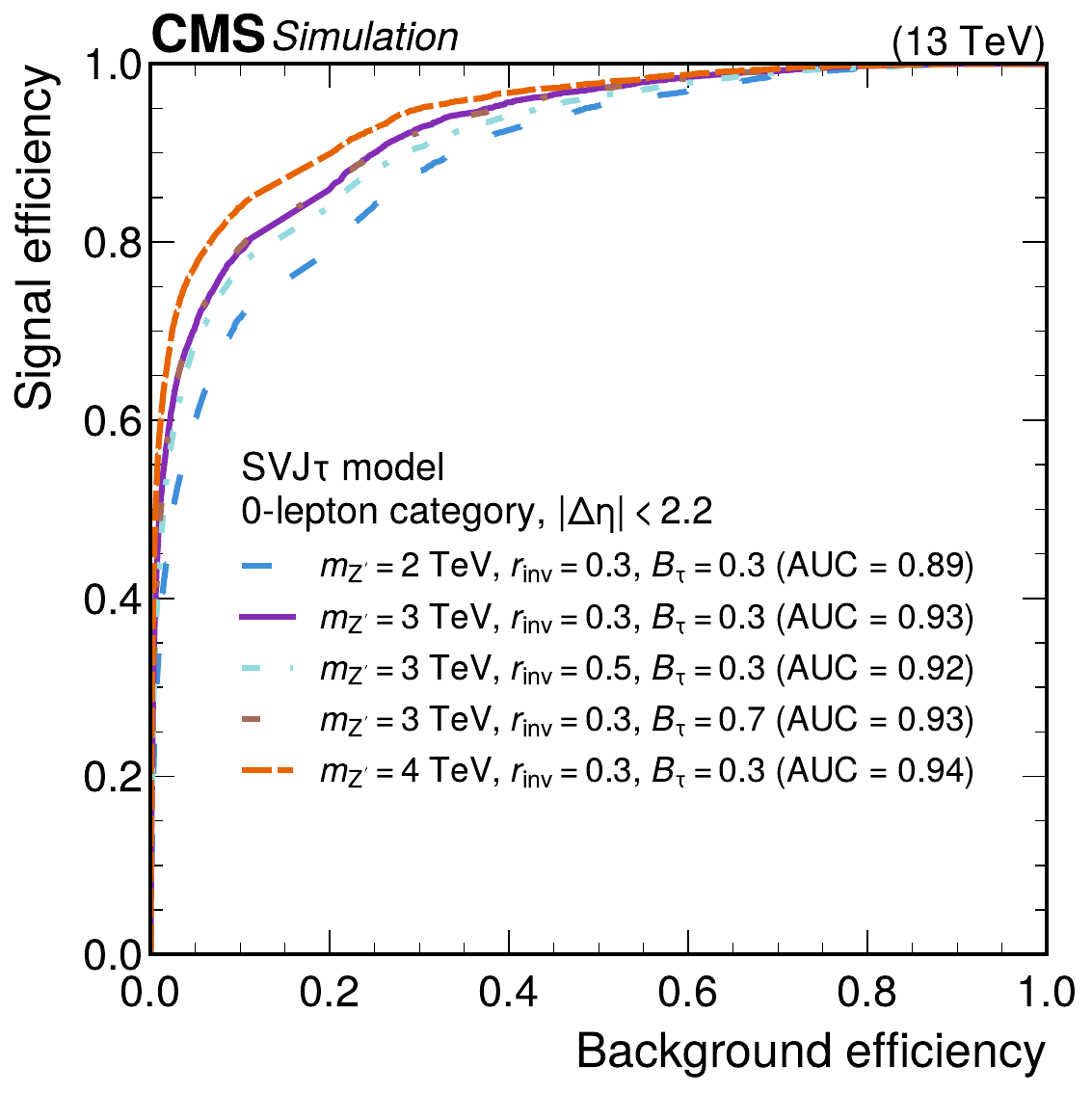}
    \includegraphics[width=0.48\linewidth]{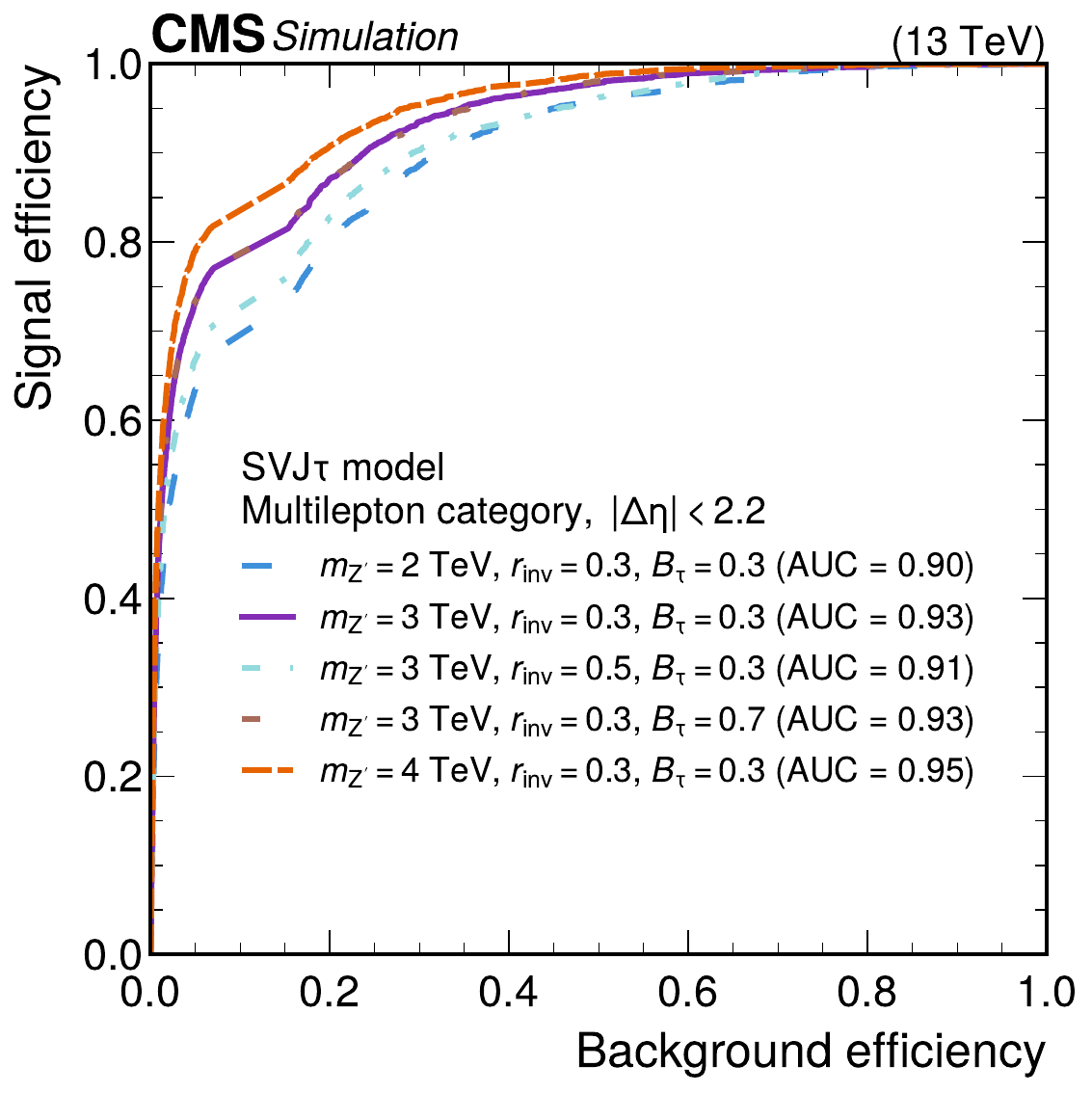}
    \caption{Left: the ROC curves presenting the possible working points of \LUNDNET in the background efficiency versus signal efficiency plane from simulations for different \svjtau signal models (with $\mDark=8\GeV$) in the 0-lepton category (\deta-extended region). Right: the ROC curves presenting the possible working points of \LUNDNET in the background efficiency versus signal efficiency plane from simulations for different \svjtau signal models (with $\mDark=8\GeV$) in the multilepton category (\deta-extended region). The AUC is computed as the area under the ROC curve for a given signal model against the total background.}
    \label{fig:lundet_performance_svjtau}
\end{figure*}

The ROCs for the two \LUNDNET models trained to identify the \svjtau signals are shown in the left and right panels of Fig.~\ref{fig:lundet_performance_svjtau} for the 0-lepton and multilepton categories, respectively. The taggers overall show high AUC values for both the \svjl and \svjtau signal models. For all three trained \LUNDNET models, higher performance is observed at larger \mZprime values and lower \rinv, while the performance has been shown to be uniform in \brtau and \mDark.

\section{Background estimation}\label{sec:bkg}

As described in Section~\ref{sec:evtsel}, in the multilepton region, the dominant backgrounds arise from electroweak processes such as \wjets and \ttbar, whereas in the 0-lepton category the main background originates from QCD multijet events. Since the background composition differs significantly between these two categories, independent background estimation techniques are required for each.

{\tolerance=1200
We employ a method, based on control samples in data, to estimate the overall background in each category of the analysis based on the \ABCDISCOTEC (ABCD with distance correlation training enhanced with closure) technique~\cite{CMS:2025cvw}. 
This technique uses a ML approach to engineer two discriminant variables that further separate the signal from the background and provide an optimal plane for the application of the ABCD method to estimate the total background in the SR.
The ABCD method relies on the assumption that two variables are independent for the background such that their plane can be divided into 4 regions, A, B, C, and D, with the event yield in each region related to the others by an efficiency ratio. The SR is defined as A, and the expected background can be estimated from the signal-depleted regions B, C, and D as: $\NAa = \NB \NC / \ND$.
In this search, the ABCD method is applied to estimate the overall background in each bin of the \mt distribution.\par}

\subsection{\ABCDISCOTEC setup}

{\tolerance=1200
To satisfy the requirements of the ABCD method, a pair of neural networks is trained to be independent of each other for the background.
The outputs of these networks, denoted as \Done and \Dtwo, are designed to be statistically independent, resulting in two decorrelated discriminators that separate signal from background. This independence is enforced by minimizing not only the binary cross-entropy (BCE) loss function \Lbce, which drives the classification of signal and background events, but also the distance correlation loss $\Ldisco(\Done, \Dtwo)$, which penalizes correlations between \Done and \Dtwo based on the distance correlation (DisCo) metric~\cite{Gabor:2007}. DisCo is more sensitive to nonlinear correlations than the standard Pearson correlation coefficient~\cite{Pearson}. Consequently, it provides a powerful metric for variables such as DNN outputs, which may exhibit complex nonlinear dependencies.\par}

In addition, the network is informed about the expected agreement between the predicted \NAa value from the background estimation and the true value, called the ``closure'', by including a loss term, \Lclosure, derived from the ABCD closure relation~\cite{CMS:2025cvw}. The \Lclosure term is calculated as: $\Lclosure = [\NAa \ND - \NB \NC]/[\NAa \ND + \NB \NC]$. A sigmoid function is used to calculate the number of events in each of the four regions to preserve the differentiability of the loss function in order to be able to minimize it through gradient descent.

\begin{figure*}[htb!]
    \centering
    \includegraphics[width=0.9\linewidth]{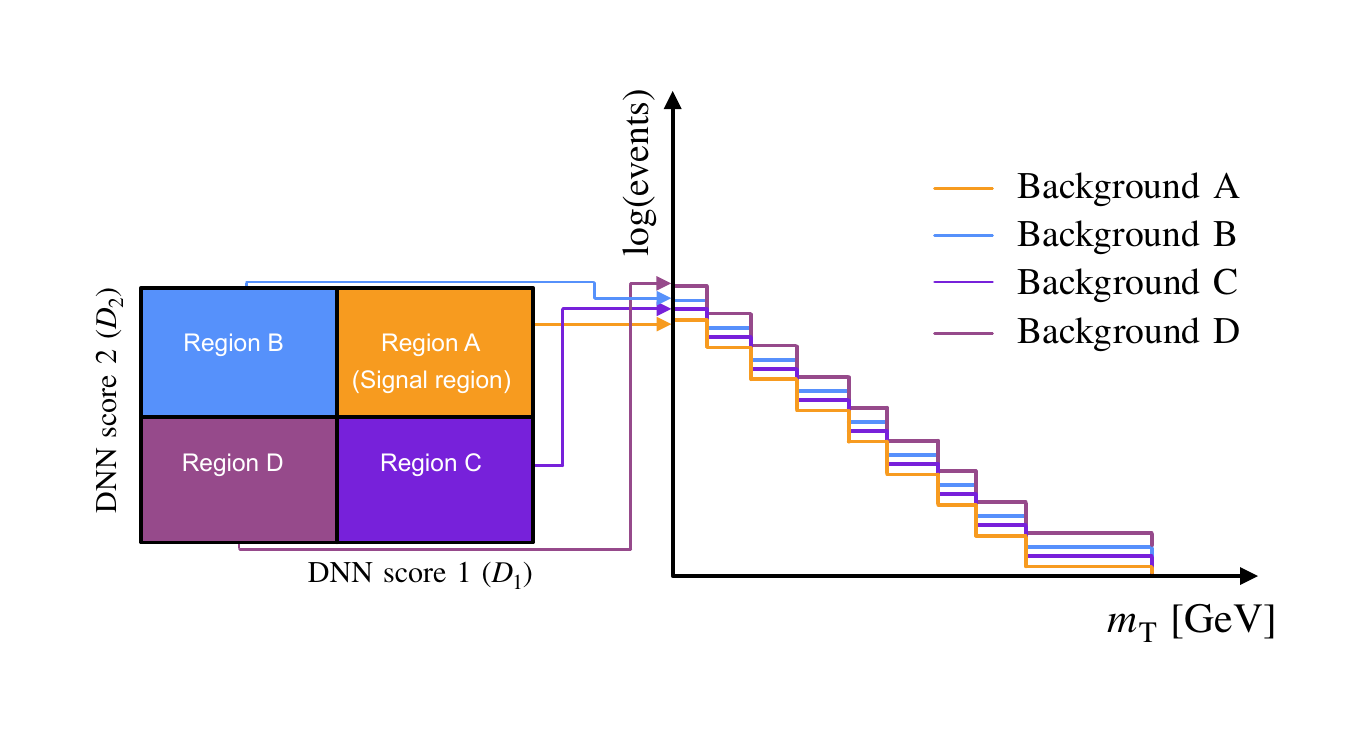}
    \caption{Sketch of the \MDABCDISCOTEC background estimation method. On the left, the ABCD plane is shown, defined by the scores of the \MDABCDISCOTEC neural network. On the right, the effect of mass decorrelation during the network training is illustrated: it results in similar \mt background distribution shapes across the different regions of the ABCD plane, enabling robust background estimation. The shift between the distributions in the sketch is due to the total normalization difference in the four regions of the ABCD plane.
    }
    \label{fig:bkg-estimation-sketch}
\end{figure*}

{\tolerance=1200
Finally, since our goal is to search for a localized excess in the \mt distribution, we introduce a modified version of the \ABCDISCOTEC method, which we call mass-decorrelated \ABCDISCOTEC or \MDABCDISCOTEC. In this approach, the discriminators are explicitly trained to be decorrelated from the \mt variable. This enables the application of the ABCD method in bins of \mt, allowing for a reliable estimation of the \mt background shape in the SR. 
In total, there are three distance correlation loss terms for the discriminators and the \mt observable: $\Ldisco(\Done,\Dtwo)$, $\Ldisco(\Done,\mt)$, and $\Ldisco(\Dtwo,\mt)$. The overall method employed to estimate the background is sketched in Fig.~\ref{fig:bkg-estimation-sketch}.\par}

{\tolerance=1200
Multi-objective optimization problems, such as this one, where multiple tasks are optimized simultaneously, are often challenging with classical approaches that combine loss terms via hyperparameter-weighted linear combinations. With such an approach, there is no guarantee that tuning the weight hyperparameters for each loss term will yield an optimal solution across all tasks.
The modified differential method of multipliers (MDMM)~\cite{Platt:1987} offers the most robust solution to this problem. 
The core principle of the method is to reformulate the multi-objective optimization as a constrained optimization. 
One loss term is chosen as the function to minimize, while the other loss terms are treated as constraints in the Lagrange multiplier formulation. 
In our case, the BCE loss is chosen as the primary target, while the DisCo and closure losses are treated as constraints:
\ifthenelse{\boolean{cms@external}}
{
\begin{equation}
    \label{eq:constrained_optimization}
    \begin{gathered}
        \makebox[\linewidth][l]{$\displaystyle \min_{\theta} \Lbce(\theta) \ \text{subject to:}$} \\[6pt]
        \begin{cases}
            \Ldisco(\Done,\Dtwo)(\theta) < \epsdiscosub{1} \\
            \Ldisco(\Done, \mt)(\theta) < \epsdiscosub{2} \\
            \Ldisco(\Dtwo, \mt)(\theta) < \epsdiscosub{3} \\
            \Lclosure(\theta) < \epsclosure
        \end{cases},
    \end{gathered}
\end{equation}
}
{
\begin{equation}
    \label{eq:constrained_optimization}
    \min_{\theta} \Lbce(\theta) \ \text{subject to: } \
    \begin{cases}
        \Ldisco(\Done,\Dtwo)(\theta) < \epsdiscosub{1} \\
        \Ldisco(\Done, \mt)(\theta) < \epsdiscosub{2} \\
        \Ldisco(\Dtwo, \mt)(\theta) < \epsdiscosub{3} \\
        \Lclosure(\theta) < \epsclosure
    \end{cases},
\end{equation}
}
where $\theta$ are the network weights and the $\epsilon$ variables are hyperparameters representing the values of the constraints.
During the training, the network weights are updated in the direction opposite to the gradient (gradient descent), while the Lagrange multipliers are updated towards the gradient (gradient ascent). 
This procedure ensures that the optima for the constrained minimization problem are attractors, rather than saddle points as they would otherwise be in the differential formulation.\par}

Three \MDABCDISCOTEC networks are trained separately: one for the \svjl signals, and two for the \svjtau signals in the 0-lepton and multilepton categories, respectively. In all cases MC simulated samples are used in the \deta-extended regions of the analysis. This ensures optimal performance for both signal models and an accurate background estimation in all the different analysis categories.

{\tolerance=1200
The input variables used for training the networks fall into three categories: event-level observables (e.g., \ptmiss, \ptmiss $\phi$, \RT, \deta, \mindphi), lepton-related features (e.g., multiplicities of electrons and muons), and the GNN scores from the \LUNDNET model for the two highest \pt jets. The modeling of these input features has been validated by comparing data and MC in the 0-lepton and multilepton categories within the \deta-extended regions used for training, with good agreement in the shapes of all distributions. For the training of the \MDABCDISCOTEC networks for the \svjtau signal, the \ptmiss $\phi$ variable has not been employed as input feature to improve the agreement between real and simulated data in presence of the large QCD background.
The electron and muon multiplicity based on the mini-isolation requirement ($\Imini(\Pell)<0.4$) is used in the \MDABCDISCOTEC network trained for the \svjl signal. In contrast, for the \svjtau signal, both in the 0-lepton and multilepton categories, the lepton multiplicities are counted using the inter-isolation variable \Iinter, introduced in Ref.~\cite{Cazzaniga:2022hxl}. The isolation \Iinter is defined as:
\begin{equation}\label{eq:inter_iso}
\Iinter(\Pell) =\frac{1}{{{\ptsub{\Pell}}} } {\sum_{\DR < 0.3} \ptsub{\widetilde{\Pell}}},   
\end{equation}
quantifying how much each lepton \Pell is isolated with respect to all the other leptons $\widetilde{\Pell}$ within a cone of radius $0.3$. After applying all the selections in Table~\ref{tab:sel}, the multiplicities for electrons and muons using $\Iinter(\Pell)$ are counted requiring $\Iinter(\Pell)>0.1$. Such features provide better discriminating power than the number of isolated leptons for the \svjtau signal. This difference arises from the softer \pt spectrum of electrons and muons characteristic of the \svjtau signal relative to \svjl.\par}

\subsection{\ABCDISCOTEC training}

We implement the \MDABCDISCOTEC method using \PYTORCH~2.0.1~\cite{Paszke:2019}. Each of the two neural networks, trained simultaneously, is a fully connected architecture with an input layer followed by 8 hidden layers of dimensions $(512,512)$, $(512,256)$, $(256,256)$, $(256,128)$, $(128,128)$, $(128,64)$, $(64,32)$, and an output layer $(32,1)$.
These layers provide the network with sufficient connections to generate a set of variables for decorrelation of the two discriminants.
Each linear layer block is followed by a batch normalization layer and a rectified linear unit (ReLU) activation function. The last layer is followed by a sigmoid activation function to ensure that the output is between 0 and 1.

Each pair of neural networks with such architecture is trained simultaneously to minimise the BCE loss and meet the imposed constraints.
The hyperparameters \epsdiscosub{1}, \epsdiscosub{2}, \epsdiscosub{3}, and \epsclosure are the values of the constraints used in the MDMM method.
The hyperparameters are tuned to minimize \Lbce, ensuring the network can decorrelate the two discriminants from each other and from the transverse mass observable and can enforce the closure relation. 
Each one of these hyperparameters has an associated damping factor hyperparameter $c_i$, which influences the rate of convergence of the loss term to the constraint.
We note that the optimal values of these hyperparameters are highly dependent on some of the other hyperparameters of the network, such as the learning rate and the batch size. 

\begin{figure*}[htb!]
    \centering
    \includegraphics[height=0.45\linewidth]{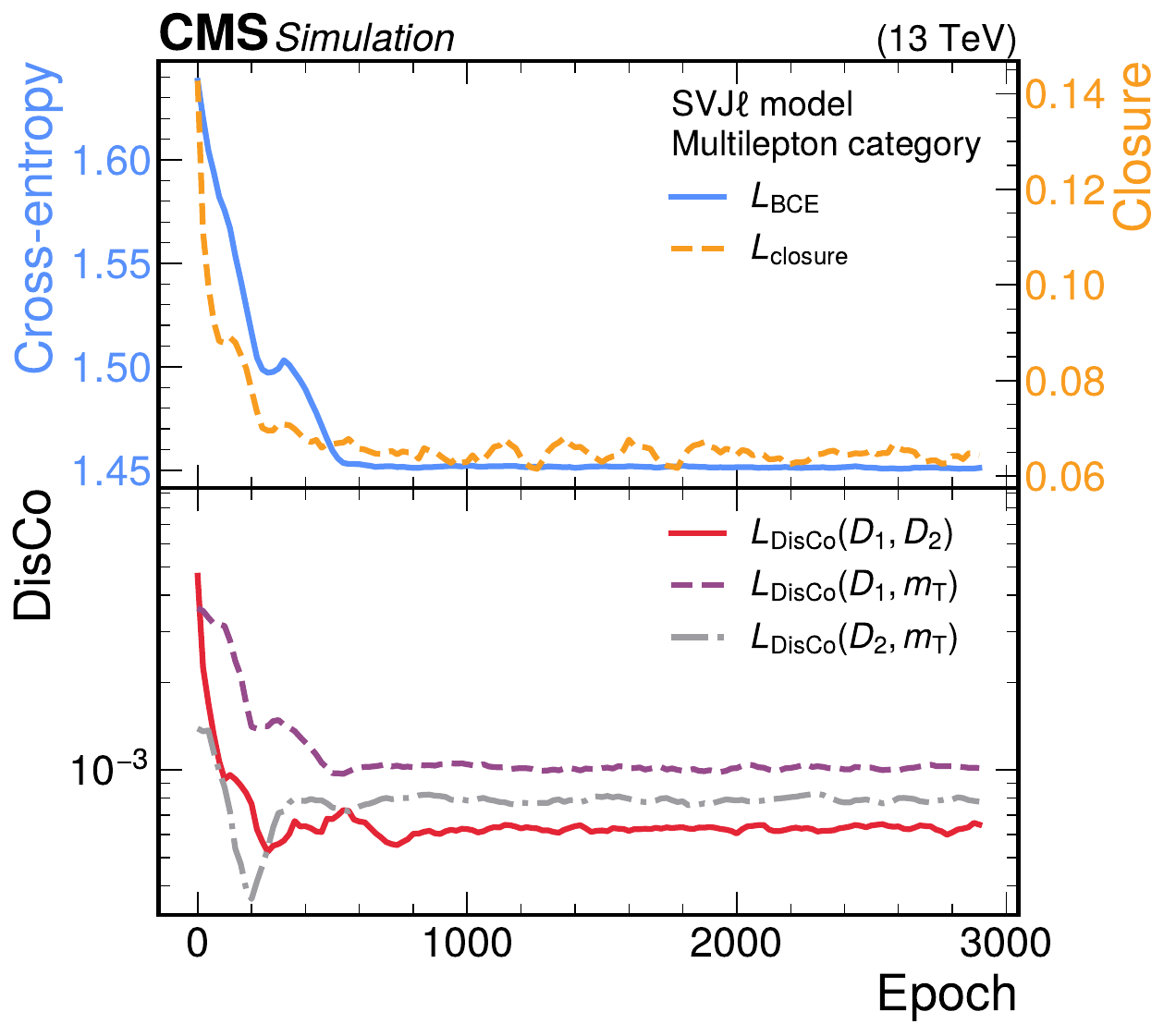}
    \includegraphics[height=0.45\linewidth]{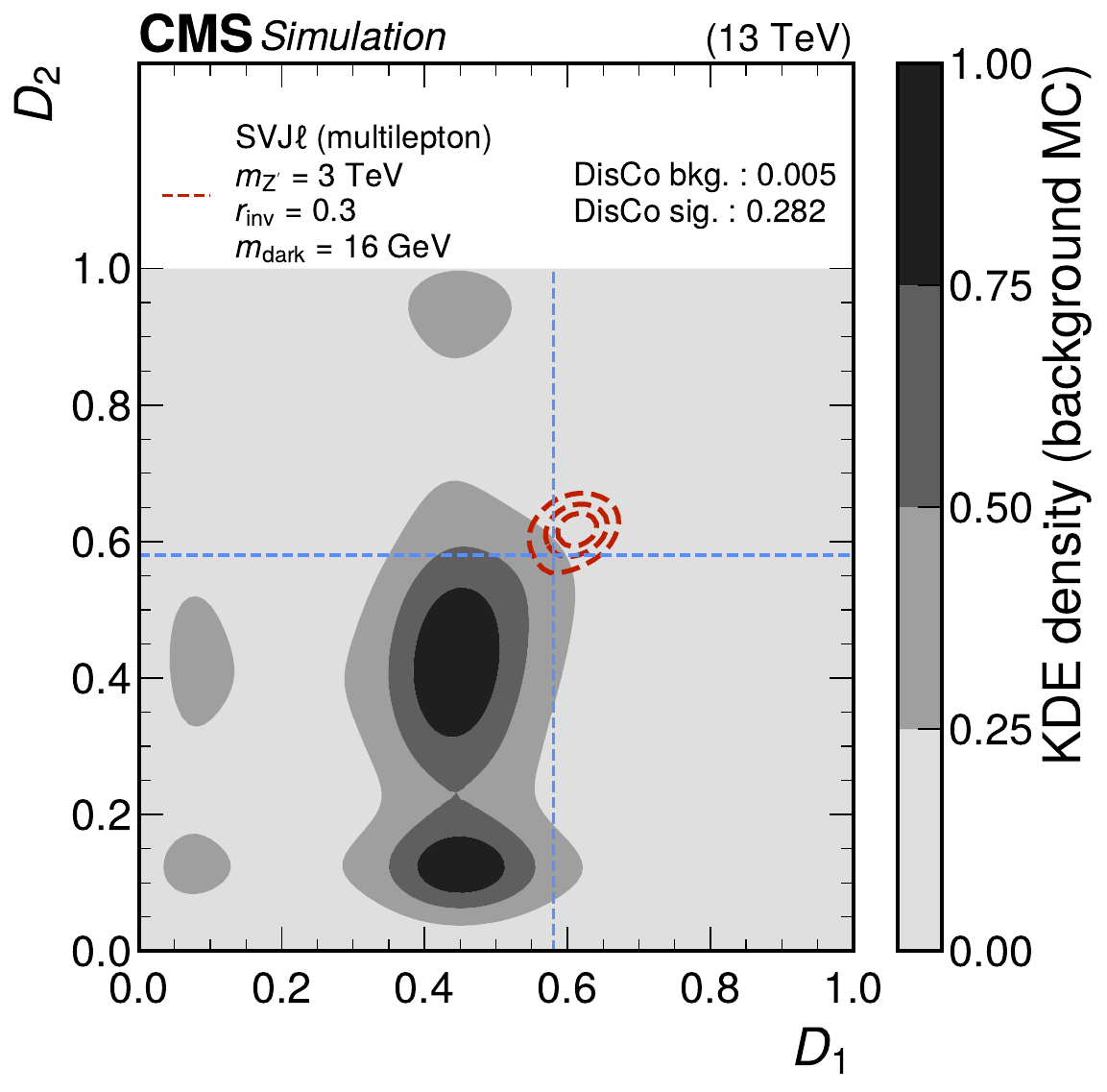}
    \caption{Left: evolution of the different components of the loss function in the training of the \MDABCDISCOTEC model for the \svjl signal in the multilepton category (\deta-extended region). Right: density distribution of simulated background and \svjl signal events, represented as Gaussian kernel density estimators (KDEs), in the ABCD plane defined by the two network scores in the multilepton category (\deta-extended region). Contour lines for the signal at 0.25, 0.5, and 0.75 are overlaid with dashed red lines. The dashed blue lines represent the ABCD boundaries chosen via the optimization procedure. In the legend the values of the DisCo for signal and background are reported.}
    \label{fig:abcd_losses_plane_svjl}
\end{figure*}

The trainings are performed with large batches, ranging from 8192 to 12\,288, depending on the specific \MDABCDISCOTEC network. Such a large batch size ensures that each batch is representative of the overall training sample, which is especially important for the DisCo and closure losses. Larger batches are employed for the trainings of the \MDABCDISCOTEC networks targeting the \svjtau signal, due to the additional parameters in the signal model, namely \brtau.
For all trainings, a learning rate scheduler is used to reduce the initial learning rate of 0.001 by a factor of 0.1 when the validation loss plateaus.
For each training, a mixture of signal hypotheses from the scans described in Section~\ref{subsec:sng_samples} is considered, depending on the signal. Each signal sample point is weighted equally.
The trainings are performed by setting the total signal and background to have the same weights. All the backgrounds are used for the training of the network, using the physical proportions of the different backgrounds.
Each of the three \MDABCDISCOTEC networks is trained by minimizing Eq.~(\ref{eq:constrained_optimization}). An example of the evolution of the different terms in the loss function that the model is trained to minimize is shown in the left panel of Fig.~\ref{fig:abcd_losses_plane_svjl} for the \MDABCDISCOTEC model trained for the \svjl signal in the multilepton category.

\begin{figure*}[htb!]
    \centering
    \includegraphics[height=0.46\linewidth]{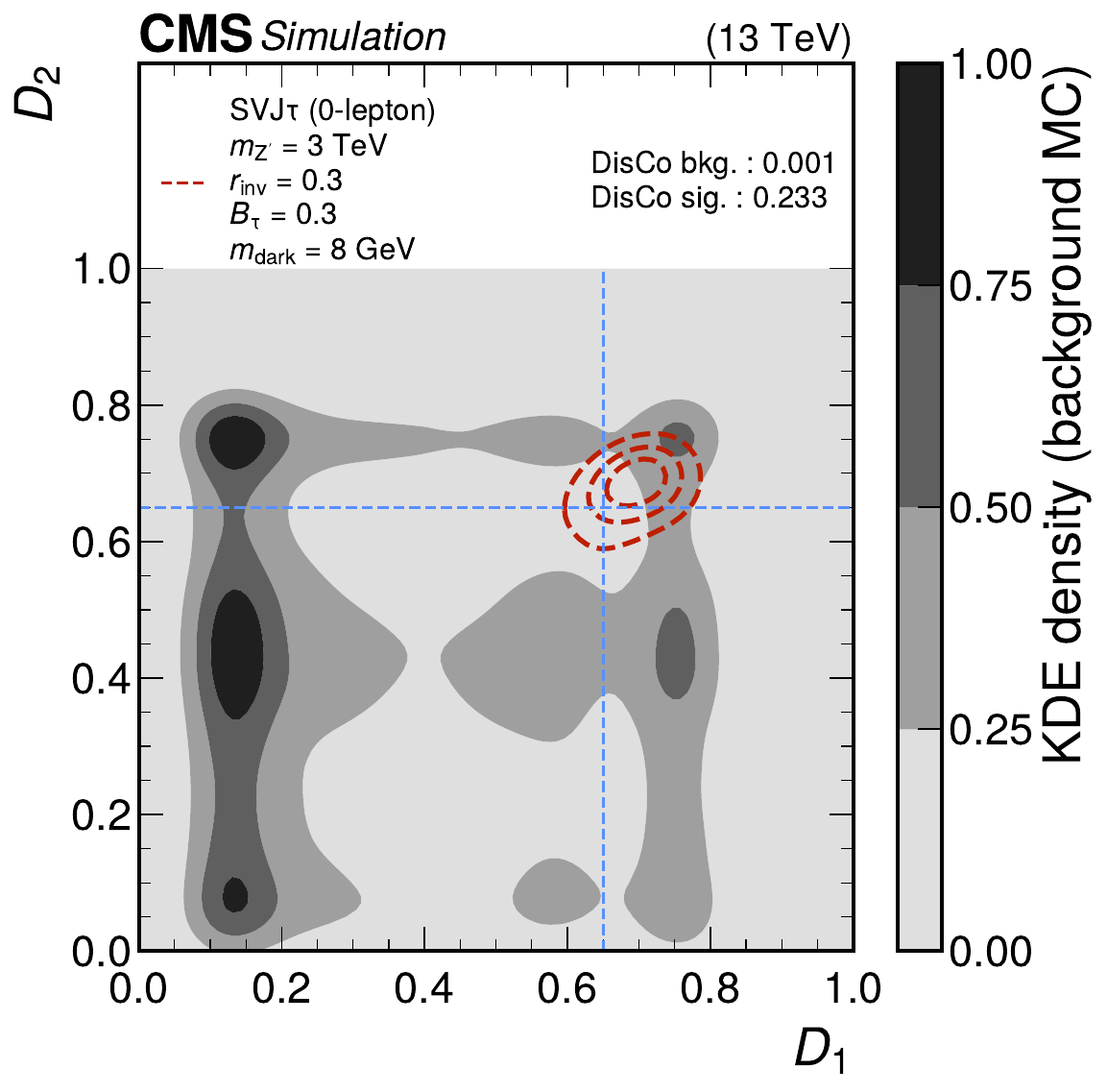}
    \includegraphics[height=0.46\linewidth]{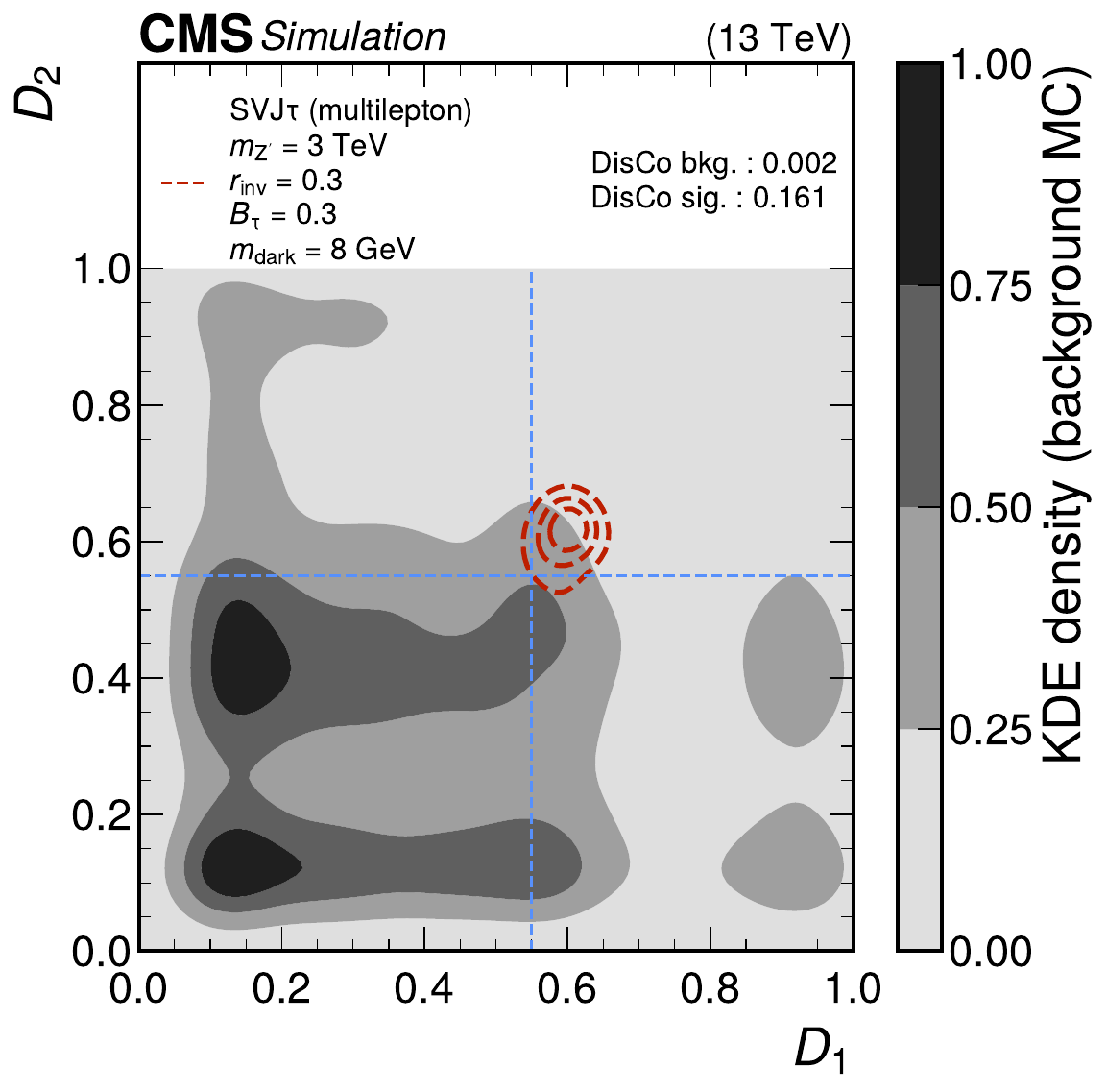}
    \caption{Left: density distribution of simulated background and \svjtau signal events, represented as Gaussian kernel density estimators (KDEs), in the ABCD plane defined by the two network scores in the 0-lepton category (\deta-extended region). Right: density distribution of simulated background and \svjtau signal events, represented as Gaussian kernel density estimators (KDEs), in the ABCD plane defined by the two network scores in the multilepton category (\deta-extended region). Contour lines for the signal at 0.25, 0.5, and 0.75 are overlaid with dashed red lines. The dashed blue lines represent the ABCD boundaries chosen via the optimization procedure. In the legends the values of the DisCo for signal and background are reported.}
    \label{fig:abcd_losses_plane_svjtau}
\end{figure*}

\subsection{\ABCDISCOTEC validation and performance}

{\tolerance=2400 A grid search is performed in each of the two-dimensional planes defined by the two discriminators. The best boundaries are found by maximizing or minimizing several figures of merit. The signal significance $S/\sqrt{S + B(1+ \nonclosure^{2})}$ in the SR is maximized to achieve the best sensitivity, where $S$ and $B$ are the expected numbers of signal and background events, and $\nonclosure$ is the expected relative difference between the simulated background events and the prediction from the ABCD method applied to background simulations. The normalized signal contamination defined in Ref.~\cite{Kasieczka:2020pil} as $(N_{\mathrm{A},S}/N_{\mathrm{A},B})^{-1}(N_{\mathrm{B},S}/N_{\mathrm{B},B} + N_{\mathrm{C},S}/N_{\mathrm{C},B} - N_{\mathrm{D},S}/N_{\mathrm{D},B})$, where $N_{\mathrm{i},S}$ and $N_{\mathrm{i},B}$ represent the expected numbers of events from signal and background simulations in region $\mathrm{i}$, is minimized by requiring that the signal fractions in each control region (CR) are smaller than the one in the SR. The quantity $\nonclosure$ is minimized to require good agreement between the background normalization predicted by the ABCD method and the expected simulated background. Finally, the $\chi^2$ between the shape predicted from the ABCD method and the simulated background is minimized to require good agreement in the different bins of \mt, where $\chi^2 = \sum_{i \in \mathrm{\mt~bin}} (N_{\mathrm{A},\mathrm{pred}}^{i} - N_{\mathrm{A},B}^{i})^2/\sigma_i^2,$
with $N_{\mathrm{A},\mathrm{pred}}^{i}$ and $N_{\mathrm{A},B}^{i}$ denoting the predicted and simulated background yields in bin $i$ of \mt, respectively, and $\sigma_i$ the corresponding uncertainty.\par}

The ABCD plane associated with the \svjl \MDABCDISCOTEC network, together with the optimized boundaries $(0.58, 0.58)$ obtained from the optimization procedure, is shown in the right panel of Fig.~\ref{fig:abcd_losses_plane_svjl} for the simulated signal and background processes. With this choice of boundaries, roughly 95\% of the remaining background is removed from region A, while approximately 55\% of the BP-\svjl model is retained.

The left panel of Fig.~\ref{fig:abcd_losses_plane_svjtau} shows the ABCD plane for the simulated signal and background processes defined by the \svjtau \MDABCDISCOTEC network in the 0-lepton category, along with the optimized boundaries $(0.65, 0.65)$. In this case as well, around 95\% of the remaining background is rejected in region A, while approximately 55\% of the BP-\svjtau signal is retained.

The right panel of Fig.~\ref{fig:abcd_losses_plane_svjtau} shows the ABCD plane for the simulated signal and background processes defined by the \svjtau \MDABCDISCOTEC network in the multilepton category, together with the optimized boundaries $(0.65, 0.65)$. After applying the ABCD selection, about 90\% of the background is removed in region~A, while roughly 75\% of the BP-\svjtau signal is retained.
As shown in Figs.~\ref{fig:abcd_losses_plane_svjl}-\ref{fig:abcd_losses_plane_svjtau}, the distance correlation between \Done and \Dtwo evaluated on background events lies in the range 0.001--0.005, demonstrating an excellent level of decorrelation between the network scores, which is essential for the application of the ABCD method. In contrast, signal events display higher distance correlation values, as decorrelation is not required for signal events for the validity of the ABCD method.

The validation of the three \MDABCDISCOTEC networks is performed for the 0-lepton and multilepton categories in the high-\deta validation region (VR), which is expected to be signal-depleted, by comparing the predicted background to the observed data in region A for each \mt bin. The estimated background distribution is found to be in good agreement with the observed data in the VR, for the 2016, 2017, and 2018 data-taking periods separately. For the \svjl signal in the multilepton category, a maximum of 0.1--10\% difference in normalization is found. For the \svjtau signal, the difference in normalization is 5--15\% in the 0-lepton category and 1--10\% in the multilepton category. Such differences are taken as a systematic uncertainty in the total background normalization, as discussed in Section~\ref{sec:syst}.

\begin{figure*}[htb!]
    \centering
    \includegraphics[width=1.0\linewidth]{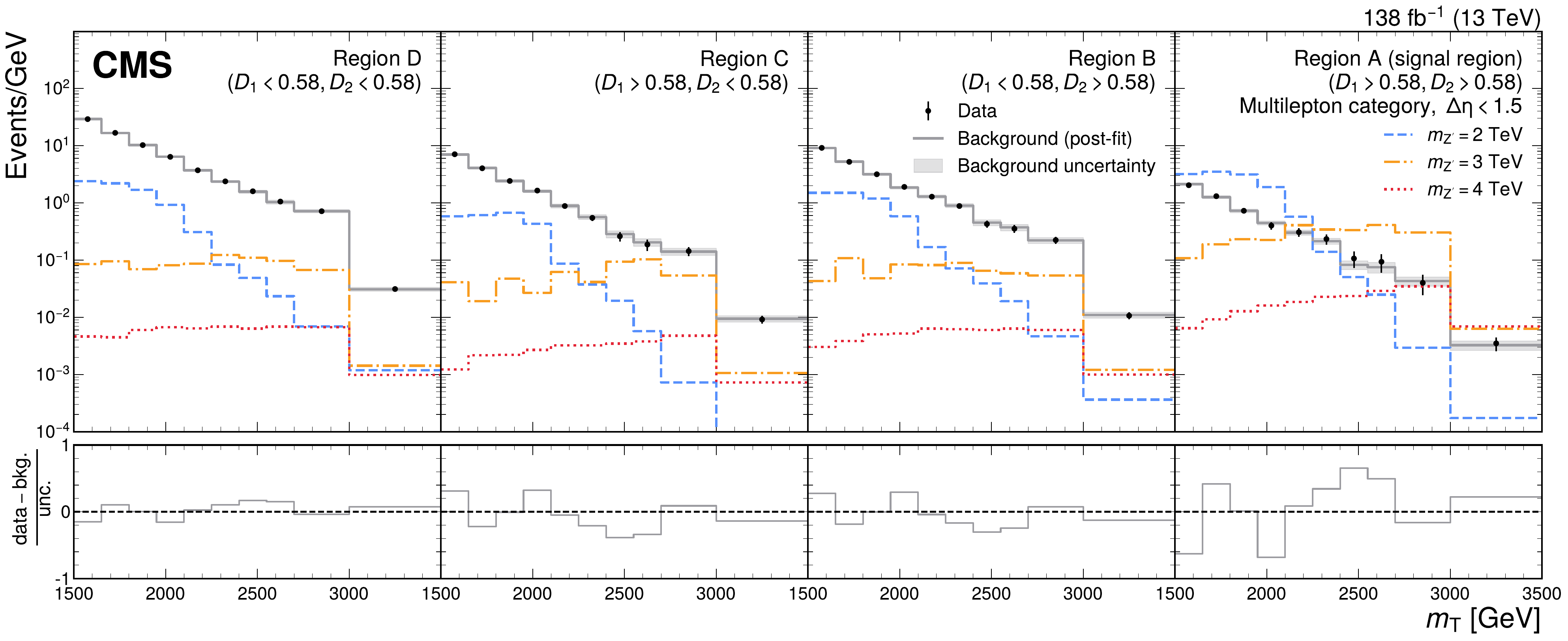}
    \caption{Comparison of estimated background and observed data in the multilepton low-\deta region for the \svjl search. The distributions from several signal model examples (with $\mDark=16\GeV$) are superimposed. The last bin of the distribution includes all events with $\mt>3\TeV$. In the upper panel, the uncertainty in the background prediction is represented by the gray bands, while statistical uncertainties on data are shown with vertical bars on the marker. In the lower panel, the difference between the data and the background prediction divided by the total uncertainty is shown.}
    \label{fig:bkg-estimation-multi-lep-svjl}
\end{figure*}

\begin{figure*}[htb!]
    \centering
    \includegraphics[width=1.0\linewidth]{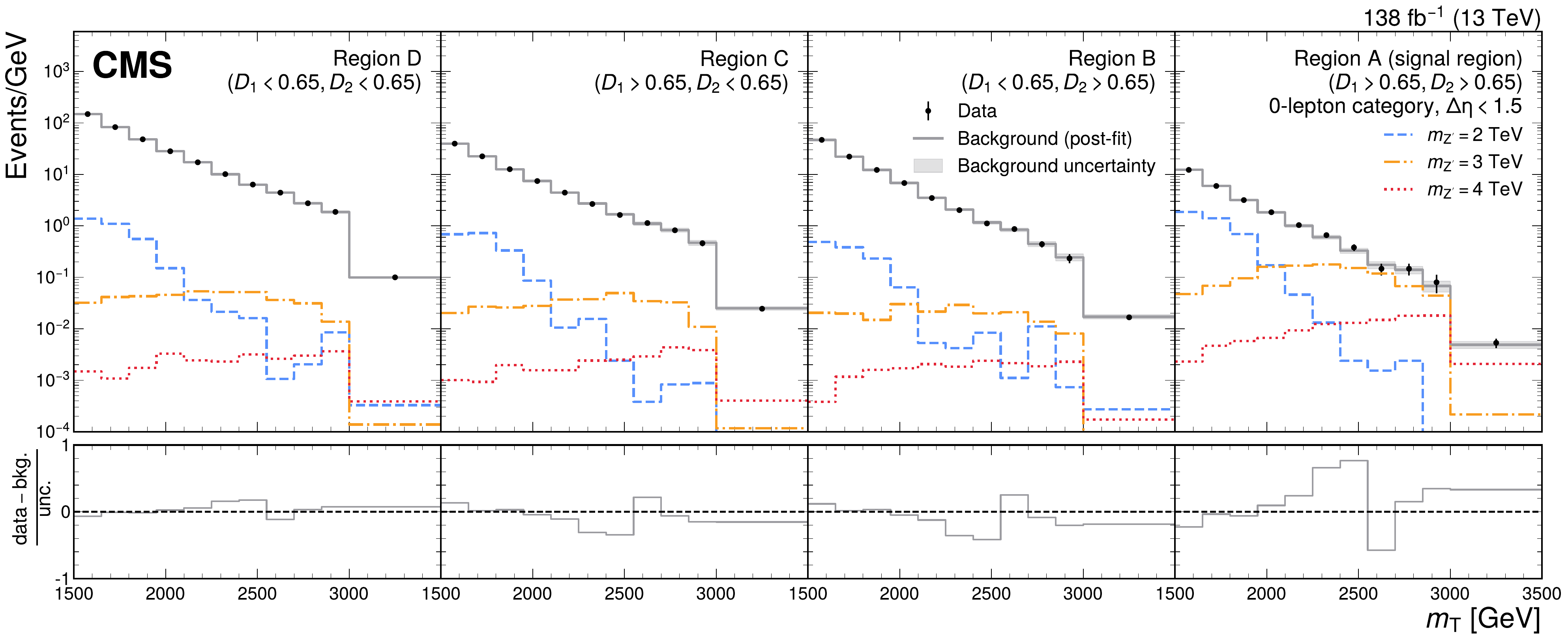}
    \caption{Comparison of estimated background and observed data in the 0-lepton low-\deta region for the \svjtau search. The distributions from several signal model examples (with $\mDark=8\GeV$) are superimposed. The last bin of the distribution includes all events with $\mt>3\TeV$. In the upper panel, the uncertainty in the background prediction is represented by the gray bands, while statistical uncertainties on data are shown with vertical bars on the marker. In the lower panel, the difference between the data and the background prediction divided by the total uncertainty is shown.}
    \label{fig:bkg-estimation-0-lep-svjtau}
\end{figure*}

\begin{figure*}[htb!]
    \centering
    \includegraphics[width=1.0\linewidth]{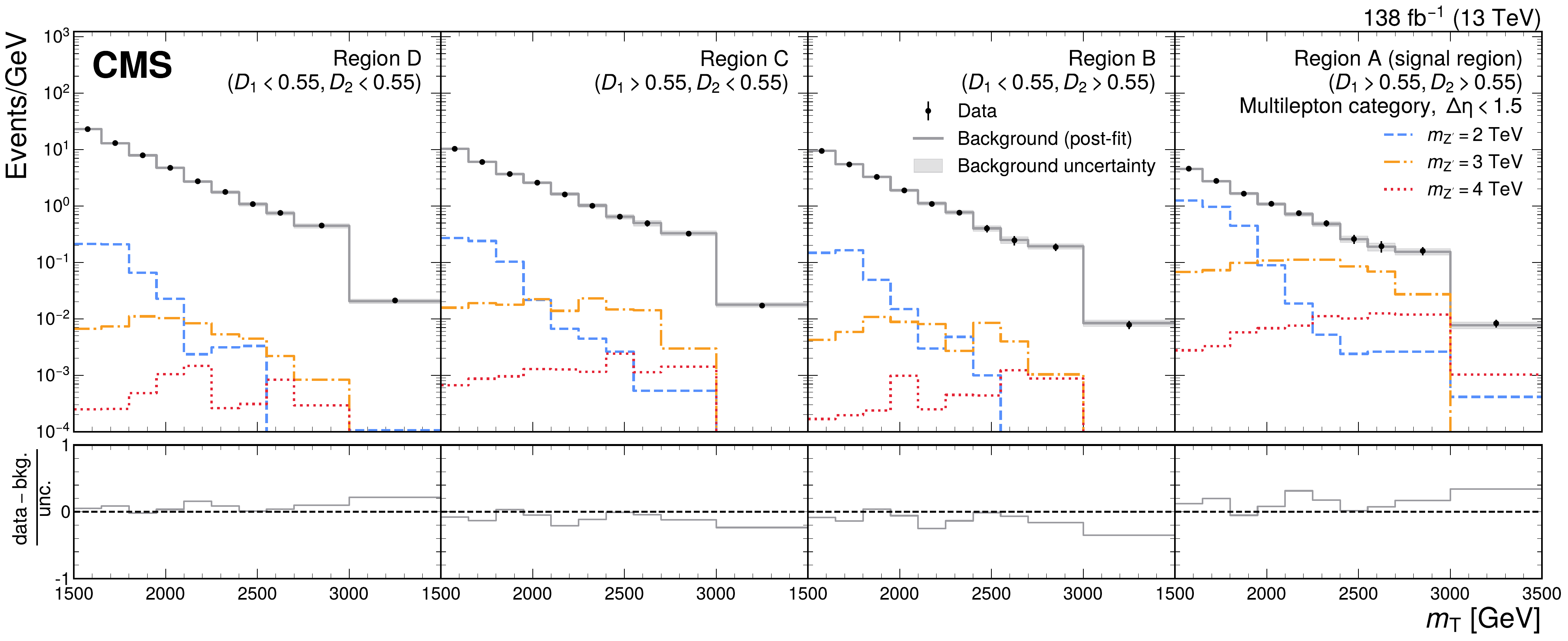}
    \caption{Comparison of estimated background and observed data in the multilepton low-\deta region for the \svjtau search. The distributions from several signal model examples (with $\mDark=8\GeV$) are superimposed. The last bin of the distribution includes all events with $\mt>3\TeV$. In the upper panel, the uncertainty on the background prediction is represented by the gray bands, while statistical uncertainties on data are shown with vertical bars on the marker. In the lower panel, the difference between the data and the background prediction divided by the total uncertainty is shown.}
    \label{fig:bkg-estimation-multi-lep-svjtau}
\end{figure*}

The results for the estimation of the background in the \mt observable for the \svjl search, obtained by applying the \MDABCDISCOTEC method in the multilepton category within the low-\deta region, are compared with the observed data in the ABCD regions in Fig.~\ref{fig:bkg-estimation-multi-lep-svjl}. For the \svjtau search, the background \mt distribution is estimated in the 0-lepton and multilepton categories within the low-\deta region; the corresponding comparisons with data in the ABCD regions are shown in Figs.~\ref{fig:bkg-estimation-0-lep-svjtau} and~\ref{fig:bkg-estimation-multi-lep-svjtau}, respectively. 
The possible contamination of the signal in the CRs of the ABCD method is taken into account by using a common signal strength modifier $\mu$ in all regions in the simultaneous fit.
A new resonance would appear as a significant excess of events in the \mt distribution in region A, and no such excess with
respect to the predictions is observed. The agreement between the background-only prediction and the data is quantified by a goodness-of-fit test based on the saturated model~\cite{bib:gof}. A p-value of 0.75 was found for the \svjl search, and a p-value of 0.98 was found for the \svjtau search. We proceed to set limits on the effective cross sections for different signal models in Section~\ref{sec:results}.

\section{Systematic uncertainties}\label{sec:syst}

The systematic uncertainties affecting the results have various sources.
We consider two types of uncertainties.
The first includes experimental uncertainties related to measurements of detector and reconstruction effects, often occurring when a correction is applied to handle some difference between data and simulation.
The second includes theory uncertainties that reflect possible variations in the generation of signal events.
Some uncertainties, called ``flat'' uncertainties, affect just the total normalization of the signal or background. Other uncertainties, called ``shape'' uncertainties, affect both the yield and the shape of the distribution.
Experimental shape uncertainties are considered to be uncorrelated between each year of data taking. In contrast, the theory uncertainties are considered to be correlated, because their sources are common across all years.
Finally, the statistical uncertainty from the limited number of simulated signal events is taken into account.
This statistical uncertainty is uncorrelated for each bin of the \MT distribution and for each year of data taking.
The effect of each systematic uncertainty on the signal yield is summarized in Table~\ref{tab:syst_effects_table}.

The flat experimental uncertainties include a 0.73\% uncertainty in the measurement of the integrated luminosity~\cite{CMS:2021xjt,CMS:2026mqq}
and a 2.0\% uncertainty in the measured trigger efficiency, to account for potential kinematic differences between the SRs and the CR used in the measurement. These uncertainties are correlated across the years.

The remaining experimental uncertainties in the signal models are shape uncertainties.
These include uncertainties in the jet energy corrections and the jet energy resolution, which are evaluated depending on the jet \pt and $\eta$,
and propagated to all jet-related kinematic variables and to the missing transverse momentum. Because the jet energy corrections are measured for SM jets, whose constituents may be different than SVJs,
we include an additional uncertainty in the jet energy scale following the same strategy as Ref.~\cite{CMS:2021dzg}, which is derived by comparing the generator-level jet \pt to the reconstructed jet \pt of simulated SVJs. This effect is also propagated to all jet-related variables and to the missing transverse momentum. The uncertainty in the electron and muon reconstruction, identification, and isolation efficiencies is applied by varying the relative scale factors by their uncertainties. The electron and muon scale factor systematic uncertainties are applied only to the categories where isolated leptons are selected (multilepton region). The uncertainty in the unclustered energy is applied by varying the unclustered energy and propagating the changes to the \ptmiss~\cite{CMS:2019ctu}. The uncertainty in the pileup reweighting is estimated by varying of the total inelastic cross section by ${\pm}5$\%~\cite{Sirunyan:2018nqx}.

The theory uncertainties are generally treated as shape uncertainties. These include the uncertainty in the PDFs, which is assessed by reweighting the generated events using the different PDF replicas~\cite{Ball:2013hta,Ball:2017nwa}; only the effect on the acceptance is considered. The uncertainty in the renormalization and factorization scales is evaluated as the envelope of all combinations obtained by varying both scales up and down by a factor of two~\cite{Kalogeropoulos:2018cke,Catani:2003zt,Cacciari:2003fi}, following the standard 7-point prescription, excluding the two extreme variations. As for the PDF uncertainty, only the effect on the acceptance is considered. The uncertainty in the parton shower model is found by similar variations of the renormalization scale used in \PYTHIA~\cite{Mrenna:2016sih};
the contributions to initial-state radiation (ISR) and final-state radiation (FSR) are considered separately.

\begin{table*}[!hbtp]
\topcaption{The range of effects on the signal yield for each signal-related systematic uncertainty in each analysis category. The variation in the yield effects arises from the different years of data taking and the range of signal models considered. Values less than 0.05\% are rounded to 0\%.}
\centering
\resizebox{\textwidth}{!}{%
\begin{tabular}{lccc}
Uncertainty & \multicolumn{3}{c}{Yield effect [\%]} \\
 & \svjl\ multilepton & \svjtau\ 0-lepton & \svjtau\ multilepton \\
\hline
Integrated luminosity & 0.73 & 0.73 & 0.73 \\ 
Trigger efficiency & 2 & 2 & 2 \\
Jet energy corrections & 0.2--10 & 0.1--18 & 0--16 \\
Jet energy resolution & 0--2 & 0--14 & 0--15\\
SVJ energy scale & 0.1--2 & 0--12 & 0--17 \\
Muon scale factors & 0.3--7 & \NA & 1.5--5\\
Electron scale factors & 2--5 & \NA & 1.2--3  \\
Unclustered energy & 0--20 & 0--20  & 0--17\\
Pileup reweighting & 0--6 & 0--7 & 0--4\\
PDF & 0--1 & 0--1  & 0--1\\
Parton shower FSR & 0--9 & 0--4 & 0--7 \\
Parton shower ISR & 0--7 & 0--4 & 0--3 \\
Renormalization and factorization scales & 0--1 & 0--1 & 0--1\\
\end{tabular}
}
\label{tab:syst_effects_table}
\end{table*}
 
Finally, we assign two systematic uncertainties to the background estimation related to different residual effects of the \MDABCDISCOTEC method.

{\tolerance=1200
A flat uncertainty in the total background yield is applied to cover possible impacts on the total normalization from nonclosure in the ABCD method, which could arise from residual correlations between the two DNN scores employed in \MDABCDISCOTEC. This uncertainty is extracted separately from each data-taking period using the observed data in the 0-lepton and multilepton high-\deta regions and applied to the SR A as an uncorrelated systematic uncertainty for each year. For the \svjl signal in the multilepton category, the total background-yield uncertainty ranges from 0.1--10\%. For the \svjtau signal, the uncertainty is 5--15\% in the 0-lepton category and 1--10\% in the multilepton category.\par}

The shape uncertainty in the background prediction has two components. 
The first component has a separate, uncorrelated uncertainty for each bin of \mt, to account for possible bin-by-bin nonclosure in the ABCD method. Due to the limited number of events in the high-\mt bins in the high-\deta region, the uncertainty is derived in the low-\deta region from the \mt distribution of simulated background events, with a Poisson uncertainty included to emulate the statistical uncertainty in observed data.
The second component of the background shape uncertainty accounts for possible variations of the \mt shape from residual correlations between the \MDABCDISCOTEC scores and \mt itself. This uncertainty is derived from the \mt distribution of the simulated background by varying the \MDABCDISCOTEC boundaries by 10\% around the nominal value and computing the relative change in the \mt shape in the SR. For each of the three planes defined by the \MDABCDISCOTEC networks, this procedure results in two separate sets of possible shape variations for the \mt distribution, one for each DNN score. For each set of shapes, an envelope is built considering the 16\% and 84\% percentiles in each \mt bin as the down and up variations, respectively.
Both shape uncertainties are derived separately for the 2016, 2017, and 2018 data-taking periods. In order to reduce the total number of nuisance parameters related to the background estimation, the two components of the shape uncertainty are added in quadrature for each \mt bin.

The uncertainty in the yield of each \mt bin for regions B, C, and D comes from the statistical uncertainty in the observed data. For region A, the uncertainty in each \mt bin is derived by propagating the statistical uncertainties from regions B, C, and D. The normalization and background shape uncertainties previously discussed are further added to the SR background prediction. 

For both the \svjl and \svjtau searches, the statistical uncertainty is the dominant contribution in the measurement of the signal strength $\mu$, accounting for about 80--90\% and 80--85\% of the total uncertainty, respectively. The total uncertainty is obtained by adding the statistical and systematic components in quadrature. The systematic contribution, dominated by the uncertainty in the background estimate, is subdominant, contributing approximately 40--60\% of the total uncertainty for the \svjl search and 50--60\% for the \svjtau search.

The 68\% confidence intervals on $\mu$ have been derived using the profile likelihood ratio test statistic~\cite{Cowan:2010js} following the procedure described in Sec. 3.2 of Ref.~\cite{Khachatryan:2014jba}. Following the asymptotic approximation, the 68\% confidence interval is defined as the set of parameter values satisfying $2\Delta \ln\mathcal{L} < 1$, where $\Delta \ln\mathcal{L}$ is the difference between the negative log-likelihood at a given value of $\mu$ and its minimum value.

\section{Results}\label{sec:results}

{\tolerance=1200
We set limits on the effective cross section $\sigmaZprime\Bdark$ at 95\% confidence level (\CL) using the modified frequentist \CLs approach~\cite{Junk:1999kv,Read:2002hq}.
The signal systematic uncertainties described in Section~\ref{sec:syst} are included in the maximum likelihood fit as nuisance parameters,
with the normalization uncertainties given log-normal prior distributions and the shape uncertainties given Gaussian prior distributions. The background normalization systematic uncertainties described in Section~\ref{sec:syst} are modeled via log-normal prior distributions, and shape background systematic uncertainties via asymmetric log-normal constraints. The fitting and limit-setting procedure is performed using the CMS \textsc{Combine} tool~\cite{CMS:2024onh}.\par}

{\tolerance=1200
For the \svjl search, the final inference is performed using the multilepton low-\deta region, whereas for the \svjtau search both the $0$-lepton and multilepton low-\deta regions are included. A common signal strength modifier is used in the simultaneous fit of the A, B, C, and D regions.
The expected limits are derived from the Asimov data set~\cite{Cowan:2010js}, which is created by replacing the observed data with the central values of the background predictions from the \MDABCDISCOTEC method when comparing the likelihoods of the background-only and signal-plus-background hypotheses.
The likelihood distributions from the Asimov data set and the observed data are computed as functions of the effective cross section
and are used to calculate the \CLs criterion following the asymptotic approximation~\cite{Cowan:2010js}.\par}

Models are excluded if the observed limit is below the product of the theoretical cross section for \PZprime boson production and the \PZprime branching fraction to dark quarks.
The results can be reinterpreted for different values of the \PZprime couplings and branching fractions in the range of validity of the narrow width approximation, ${\wZprime/\mZprime \lesssim 10\%}$~\cite{Boveia:2016mrp}.

\begin{figure*}[htb!]
\centering
\includegraphics[width=0.49\linewidth]{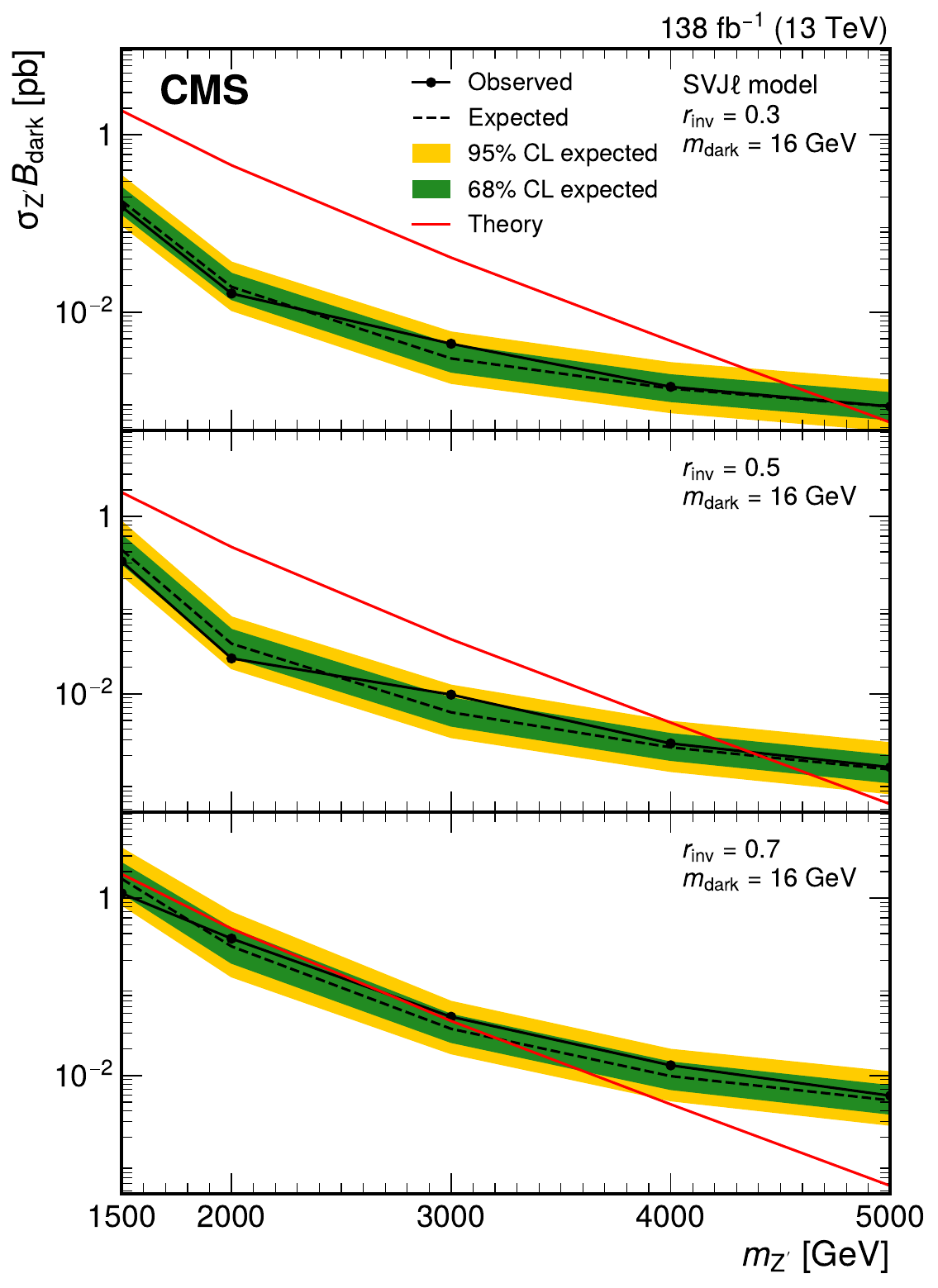}
\includegraphics[width=0.49\linewidth]{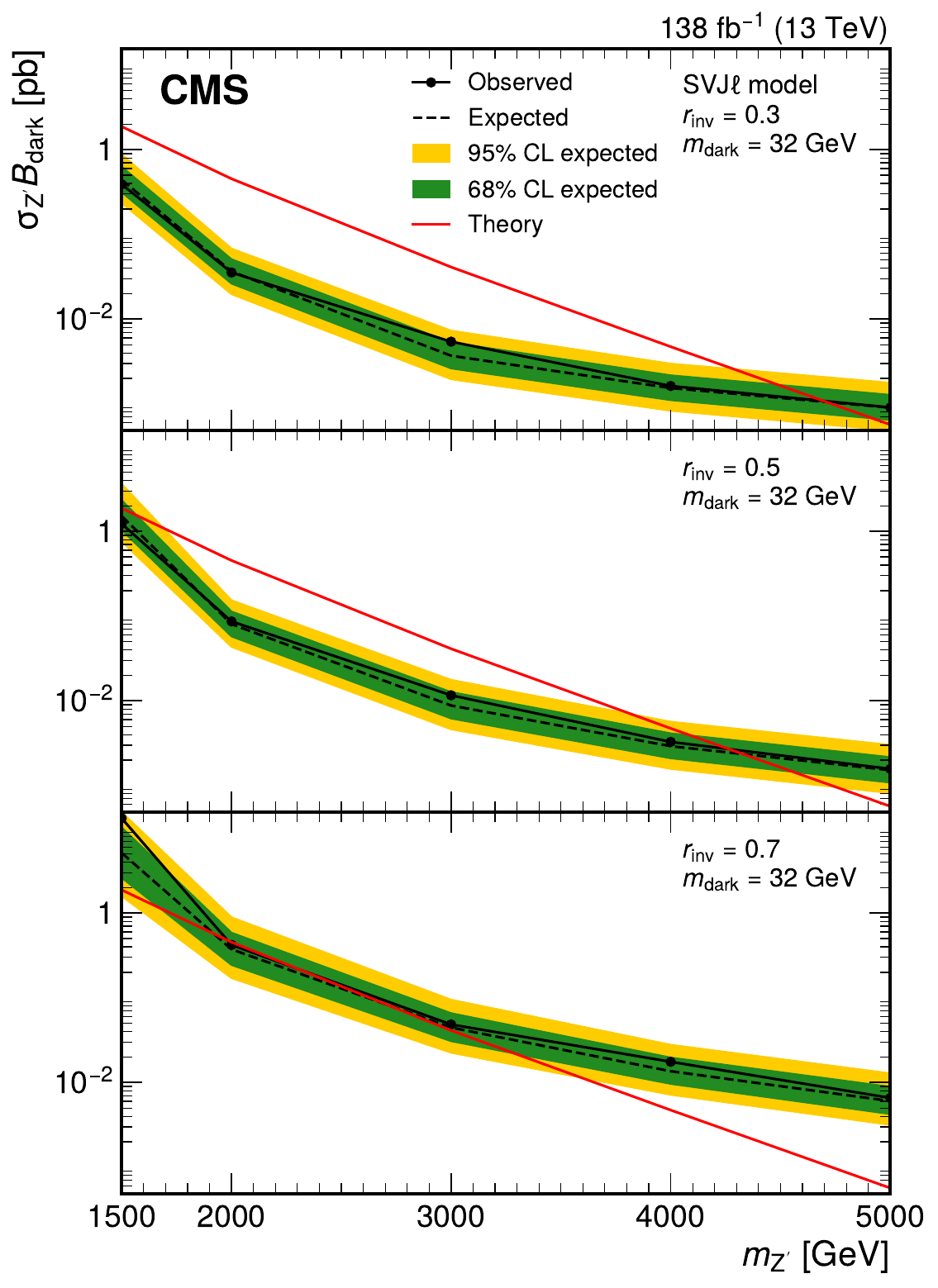}
\caption{The 95\% \CL upper limits on $\sigmaZprime\Bdark$ for the \svjl model as functions of \mZprime, for \rinv values of 0.3 (upper), 0.5 (middle), and 0.7 (lower), and $\mDark=16$ (left) and 32\GeV (right).
The red solid line labeled ``Theory'' represents the product of the nominal \PZprime cross section and \Bdark.
}
\label{fig:limit_svjl}
\end{figure*}

The expected and observed limits on $\sigmaZprime\Bdark$ for the \svjl search are shown in Fig.~\ref{fig:limit_svjl} as functions of \mZprime for different values of \rinv and \mDark. The expected (observed) 95\% \CL upper limits from the inclusive regions exclude ranges of up to $1.5 < \mZprime < 4.7\TeV$ ($1.5 < \mZprime < 4.7\TeV$) for $\rinv=0.3$ and $\mDark=16\GeV$. The widest exclusion range is found for moderate \rinv values, for which the selection efficiency is higher, the jet tagger is more powerful, and consequently the background rejection of the \MDABCDISCOTEC is larger.
For signal models with high \rinv values, the efficiency of the selection in this search decreases and both the jet- and event-level discriminators have worse performance, limiting the sensitivity.

\begin{figure*}[htb!]
\centering
\includegraphics[width=0.49\linewidth]{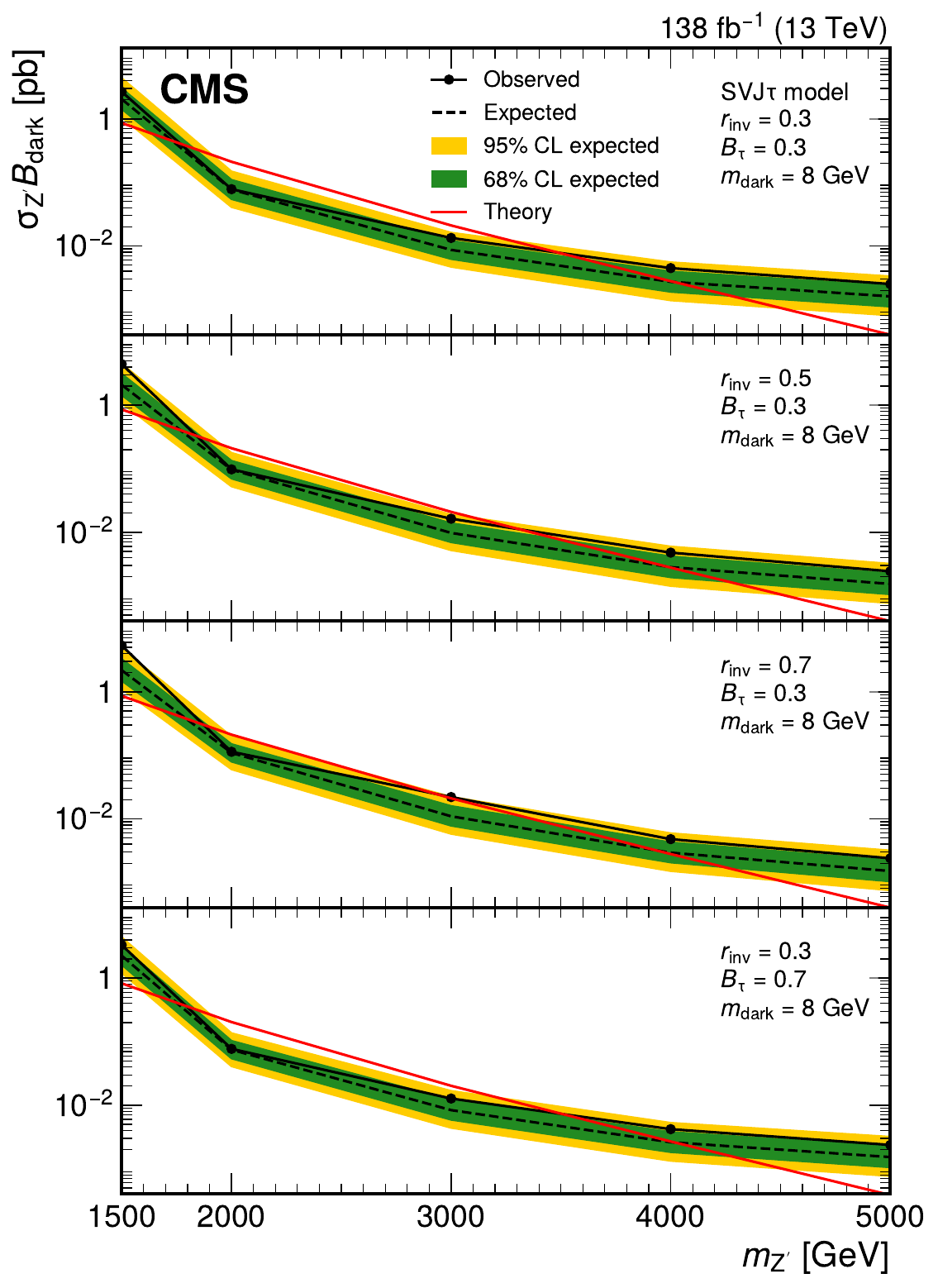}
\includegraphics[width=0.49\linewidth]{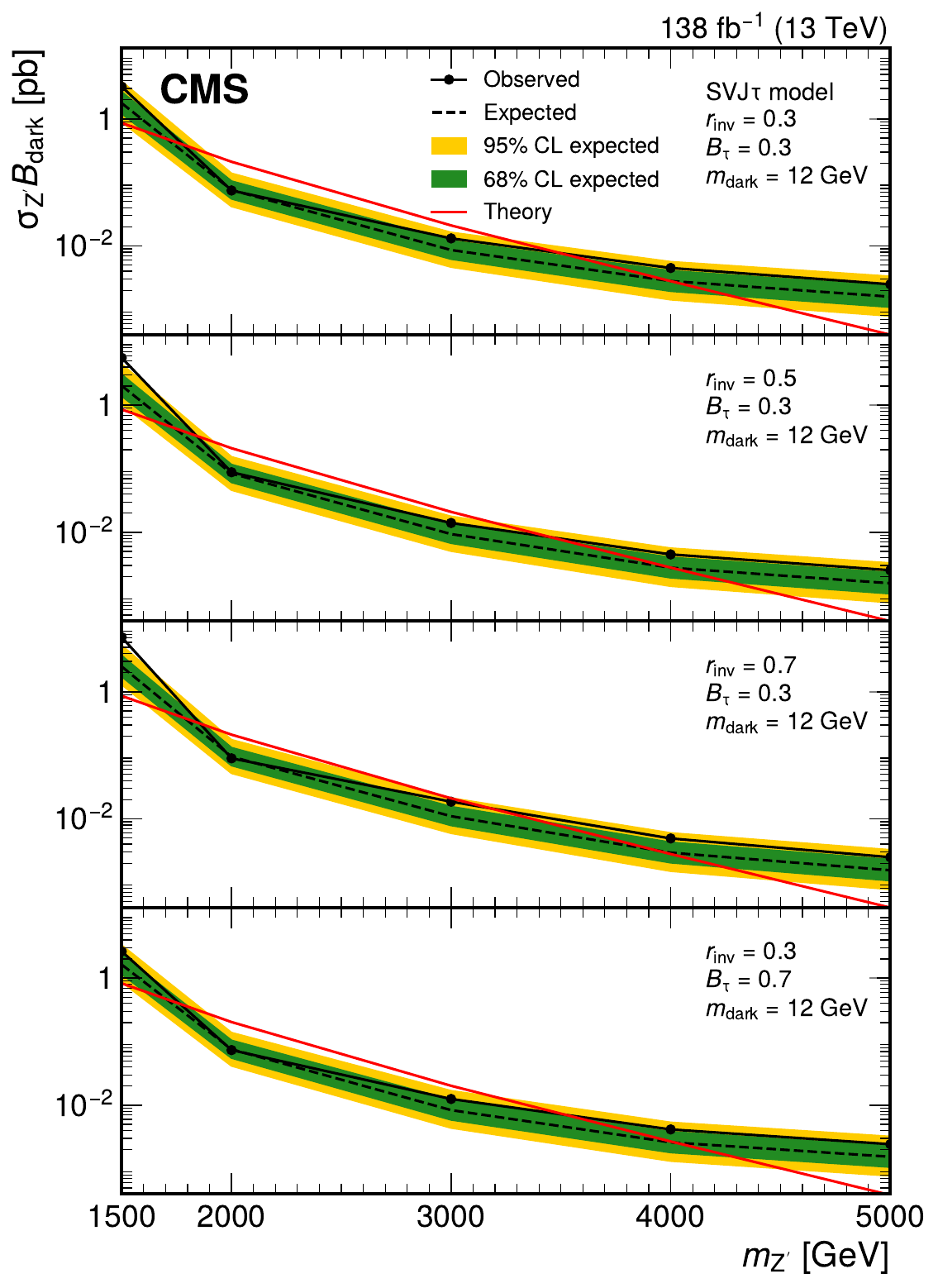}
\caption{The 95\% \CLs upper limits on $\sigmaZprime\Bdark$ for the \svjtau model as functions of \mZprime, for $\rinv=0.3$ ($\brtau=0.3$) (upper); $\rinv=0.5$ ($\brtau=0.3$) (upper middle);  $\rinv=0.7$ ($\brtau=0.3$) (lower middle); $\rinv=0.3$ ($\brtau=0.7$) (lower); and $\mDark=8$ (left) and $12\GeV$ (right). The red solid line labeled ``Theory'' represents the product of the nominal \PZprime cross section and \Bdark.
}
\label{fig:limit_svjtau}
\end{figure*}

The expected and observed limits on $\sigmaZprime\Bdark$ for the \svjtau search are shown in Fig.~\ref{fig:limit_svjtau} as functions of \mZprime for different values of \rinv, \brtau and \mDark. The expected (observed) 95\% CL upper limits exclude masses in the range $1.8 < \mZprime < 4\TeV$ ($1.8 < \mZprime < 3.5\TeV$) for $\rinv = 0.3$, $\brtau = 0.3$, and $\mDark = 8\GeV$. As \brtau directly affects \Bdark, which enters the cross section limit, and since the jet tagger and event tagger show similar performance across all tested \brtau values, only a very mild dependence on this parameter is expected. The broadest exclusion is obtained at moderate \rinv values across all \brtau choices, where the signal selection efficiency is higher, the jet tagger performs better, and the background rejection of the \MDABCDISCOTEC is correspondingly stronger. The 0-lepton category primarily contributes to the exclusion of high-mass signals, above 3\TeV for \mZprime, whereas the multilepton category improves the sensitivity at lower masses. This is due to the suppression of the large QCD background in the multilepton category, which predominantly populates the low-\mt\ region.

For the signal hypotheses at $\mZprime=3\TeV$ and $\rinv=0.7$, a small discrepancy between the observed and expected limits was found. The local significance of the observed departure reaches its maximum at $\mDark = 8\GeV$, with a value of 2.4 standard deviations. When computing the look-elsewhere effect~\cite{Gross:2010qma}, the mediator mass range of 1.5 to 5\TeV is considered with the scan step size of 0.5\TeV, resulting in a global significance of 1.6 standard deviations.

\section{Summary}\label{sec:summary}

{\tolerance=1200
The first search for resonant production of lepton-enriched semivisible jets has been presented. Two scenarios are considered: the first leading to semivisible jets enriched in all lepton flavors (\svjl signature), and the second leading to semivisible jets enriched in tau leptons (\svjtau signature).
The search uses proton-proton collision data collected with the CMS detector in 2016--2018, corresponding to an integrated luminosity of 138\fbinv at a center-of-mass energy of 13\TeV.
The signal models considered arise from a dark sector with multiple flavors of dark quarks that are charged under a dark confining force, giving rise to sprays of collimated stable and unstable dark hadrons.
The stable dark hadrons constitute dark matter candidates, whereas the unstable dark hadrons decay promptly to standard model quarks and leptons, producing lepton-enriched semivisible jets.\par}

In the \svjl scenario, the hidden sector communicates with the standard model via multiple portals: a \PZprime boson and a dark photon \PAprime. The \PZprime mediator has a {\TeVns}-scale mass and can decay to dark quarks, whereas the \PAprime mediator mainly governs the branching fractions for the dark hadron decays to leptons and quarks of all generations.
In the \svjtau scenario, the \PZprime boson couples to tau leptons, quarks, and dark quarks. The dark hadrons decay predominantly into the heaviest up-type quark kinematically accessible and into tau leptons.  

We adopt a machine-learning approach, employing an extension of the \LUNDNET algorithm to distinguish lepton-enriched semivisible jets from standard model jets. Additionally, we utilize a deep neural network that takes the \LUNDNET discriminators and other event-level and lepton-related variables as input to improve the discrimination of the \svjl and \svjtau signals from background and to estimate the background in the signal region. The data are found to agree with the standard model within uncertainties, and exclusion limits at 95\% confidence level on the \svjl and \svjtau models are established by scanning several hypotheses of the signal model parameters. For the \svjl (\svjtau) signature, \mZprime masses are excluded in the range 1.5--4.7 (1.8--3.5) \TeV, depending on \mDark and \rinv (\mDark, \rinv, and \brtau).

This analysis targets, for the first time, the \svjl and \svjtau final states, complementing existing searches for dijet resonances, dark matter in events with missing transverse momentum and initial-state radiation, and semivisible jets in fully hadronic final states. Compared to the existing fully hadronic semivisible jet results, the searches presented explore a new parameter space.

\begin{acknowledgments}
We congratulate our colleagues in the CERN accelerator departments for the excellent performance of the LHC and thank the technical and administrative staffs at CERN and at other CMS institutes for their contributions to the success of the CMS effort. In addition, we gratefully acknowledge the computing centres and personnel of the Worldwide LHC Computing Grid and other centres for delivering so effectively the computing infrastructure essential to our analyses. Finally, we acknowledge the enduring support for the construction and operation of the LHC, the CMS detector, and the supporting computing infrastructure provided by the following funding agencies: SC (Armenia), BMFWF and FWF (Austria); FNRS and FWO (Belgium); CNPq, CAPES, FAPERJ, FAPERGS, and FAPESP (Brazil); MES and BNSF (Bulgaria); CERN; CAS, MoST, and NSFC (China); MINCIENCIAS (Colombia); MSES and CSF (Croatia); RIF (Cyprus); SENESCYT (Ecuador); ERC PRG and PSG, TARISTU24-TK10 and MoER TK202 (Estonia); Academy of Finland, MEC, and HIP (Finland); CEA and CNRS/IN2P3 (France); SRNSF (Georgia); BMFTR, DFG, and HGF (Germany); GSRI (Greece); MATE and NKFIH (Hungary); DAE and DST (India); IPM (Iran); SFI (Ireland); INFN (Italy); MSIT and NRF (Republic of Korea); MES (Latvia); LMTLT (Lithuania); MOE and UM (Malaysia); BUAP, CINVESTAV, CONACYT, LNS, SEP, and UASLP-FAI (Mexico); MOS (Montenegro); MBIE (New Zealand); PAEC (Pakistan); MSHE, NSC, and NAWA (Poland); FCT (Portugal); MESTD (Serbia); MICIU/AEI and PCTI (Spain); MOSTR (Sri Lanka); Swiss Funding Agencies (Switzerland); MST (Taipei); MHESI (Thailand); TUBITAK and TENMAK (T\"{u}rkiye); NASU (Ukraine); STFC (United Kingdom); DOE and NSF (USA).

\begin{sloppypar}
\setlength\emergencystretch{\hsize}
\hyphenation{Rachada-pisek} Individuals have received support from the Marie-Curie programme and the European Research Council and Horizon 2020 Grant, contract Nos.\ 675440, 724704, 752730, 758316, 765710, 824093, 101115353, 101002207, 101001205, and COST Action CA16108 (European Union); the Leventis Foundation; the Alfred P.\ Sloan Foundation; the Alexander von Humboldt Foundation; the Science Committee, project no. 22rl-037 (Armenia); the Fonds pour la Formation \`a la Recherche dans l'Industrie et dans l'Agriculture (FRIA) and Fonds voor Wetenschappelijk Onderzoek contract No. 1228724N (Belgium); the Beijing Municipal Science \& Technology Commission, No. Z191100007219010, the Fundamental Research Funds for the Central Universities, the Ministry of Science and Technology of China under Grant No. 2023YFA1605804, the Natural Science Foundation of China under Grant No. 12535004, and USTC Research Funds of the Double First-Class Initiative No.\ YD2030002017 (China); the Ministry of Education, Youth and Sports (MEYS) of the Czech Republic; the Shota Rustaveli National Science Foundation (Georgia); the Deutsche Forschungsgemeinschaft (DFG), among others, under Germany's Excellence Strategy -- EXC 2121 ``Quantum Universe" -- 390833306, and under project number 400140256 - GRK2497; the Hellenic Foundation for Research and Innovation (HFRI), Project Number 2288 (Greece); the Hungarian Academy of Sciences, the New National Excellence Program - \'UNKP, the NKFIH research grants K 131991, K 138136, K 143460, K 143477, K 147557, K 146913, K 146914, K 147048, TKP2021-NKTA-64, and 2025-1.1.5-NEMZ\_KI-2025-00004, and MATE KKP and KKPCs Research Excellence and Flagship Research Groups grants (Hungary); the Council of Science and Industrial Research, India; ICSC -- National Research Centre for High Performance Computing, Big Data and Quantum Computing, FAIR -- Future Artificial Intelligence Research, and CUP I53D23001070006 (Mission 4 Component 1), funded by the NextGenerationEU program, the Italian Ministry of University and Research (MUR) under Bando PRIN 2022 -- CUP I53C24002390006, PRIN PRIMULA 2022RBYK7T (Italy); the Latvian Council of Science; the Ministry of Science and Higher Education, project no. 2022/WK/14, and the National Science Centre, contracts Opus 2021/41/B/ST2/01369, 2021/43/B/ST2/01552, 2023/49/B/ST2/03273, and the NAWA contract BPN/PPO/2021/1/00011 (Poland); the Funda\c{c}\~ao para a Ci\^encia e a Tecnologia (Portugal); the National Priorities Research Program by Qatar National Research Fund; MICIU/AEI/10.13039/501100011033, ERDF/EU, ``European Union NextGenerationEU/PRTR", projects PID2022-142604OB-C21, PID2022-139519OB-C21, PID2023-147706NB-I00, PID2023-148896NB-I00, PID2023-146983NB-I00, PID2023-147115NB-I00, PID2023-148418NB-C41, PID2023-148418NB-C42, PID2023-148418NB-C43, PID2023-148418NB-C44, PID2024-158190NB-C22, RYC2021-033305-I, RYC2024-048719-I, CNS2023-144781, CNS2024-154769 and Plan de Ciencia, Tecnolog{\'i}a e Innovaci{\'o}n de Asturias, Spain; the Chulalongkorn Academic into Its 2nd Century Project Advancement Project, the National Science, Research and Innovation Fund program IND\_FF\_68\_369\_2300\_097, and the Program Management Unit for Human Resources \& Institutional Development, Research and Innovation, grant B39G680009 (Thailand); the Eric \& Wendy Schmidt Fund for Strategic Innovation through the CERN Next Generation Triggers project under grant agreement number SIF-2023-004; the Kavli Foundation; the Nvidia Corporation; the SuperMicro Corporation; the Welch Foundation, contract C-1845; and the Weston Havens Foundation (USA).
\end{sloppypar}

\end{acknowledgments}\section*{Data availability} Release and preservation of data used by the CMS Collaboration as the basis for publications is guided by the  \href{https://doi.org/10.7483/OPENDATA.CMS.1BNU.8V1W}{CMS data preservation, re-use and open access policy}.

\bibliography{auto_generated}
\cleardoublepage \appendix\section{The CMS Collaboration \label{app:collab}}\begin{sloppypar}\hyphenpenalty=5000\widowpenalty=500\clubpenalty=5000\cmsinstitute{Yerevan Physics Institute, Yerevan, Armenia}
{\tolerance=6000
A.~Gevorgyan\cmsorcid{0000-0003-2751-9489}, A.~Hayrapetyan, V.~Makarenko\cmsorcid{0000-0002-8406-8605}, A.~Tumasyan\cmsAuthorMark{1}\cmsorcid{0009-0000-0684-6742}
\par}
\cmsinstitute{Marietta Blau Institute for Particle Physics, Vienna, Austria}
{\tolerance=6000
W.~Adam\cmsorcid{0000-0001-9099-4341}, L.~Benato\cmsorcid{0000-0001-5135-7489}, T.~Bergauer\cmsorcid{0000-0002-5786-0293}, M.~Dragicevic\cmsorcid{0000-0003-1967-6783}, S.~Gundacker\cmsorcid{0000-0003-2087-3266}, A.K.~Guven\cmsorcid{0009-0004-5670-5138}, P.S.~Hussain\cmsorcid{0000-0002-4825-5278}, M.~Jeitler\cmsAuthorMark{2}\cmsorcid{0000-0002-5141-9560}, N.~Krammer\cmsorcid{0000-0002-0548-0985}, A.~Li\cmsorcid{0000-0002-4547-116X}, D.~Liko\cmsorcid{0000-0002-3380-473X}, M.~Matthewman, A.~Pfeiffer\cmsorcid{0000-0001-5328-448X}, J.~Schieck\cmsAuthorMark{2}\cmsorcid{0000-0002-1058-8093}, R.~Sch\"{o}fbeck\cmsAuthorMark{2}\cmsorcid{0000-0002-2332-8784}, M.~Shooshtari\cmsorcid{0009-0004-8882-4887}, M.~Sonawane\cmsorcid{0000-0003-0510-7010}, N.~Van~Den~Bossche\cmsorcid{0000-0003-2973-4991}, W.~Waltenberger\cmsorcid{0000-0002-6215-7228}, C.E.~Wulz\cmsAuthorMark{2}\cmsorcid{0000-0001-9226-5812}
% ...
\par}
$^{1}$Also at Yerevan State University, Yerevan, Armenia\\
$^{2}$Also at Technische Universit\"{a}t Wien, Vienna, Austria\\
% ...
\end{sloppypar}
\end{document}